 \documentclass{aa}
\usepackage{graphicx}
\usepackage{txfonts}
\begin{document}


\title{
Warp signatures of the Galactic disk as seen in mid infrared from Midcourse 
Space Experiment}

\author{S. Vig \and S.K. Ghosh \and D.K. Ojha}

 \offprints{S.K. Ghosh, \email{swarna@tifr.res.in}}

\institute{Tata Institute of Fundamental Research, Homi Bhabha Road,
Mumbai (Bombay) 400 005, India }

\date{Accepted 11 March 2005}

\abstract{The gross features in the distribution of stars as well as warm
(T $\ga$ 100 K)
 interstellar dust in the Galactic disk have been investigated using
 the recent mid infrared survey by Midcourse Space Experiment (MSX)
 at 8, 12, 14 \& 21 $\mu$m bands. An attempt has been made to determine
 the location of the Galactic mid-plane at various longitudes, using
 two approaches : (i) fitting exponential functions to the latitude
 profiles and (ii) statistical indicators. The former method
 is successful for the inner Galaxy ($-90^{\circ} < l \le 90^{\circ}$), and 
 quantifies characteristic angular scales, $\gamma$, along latitude. These
 $\gamma$s have been translated to  linear
 scale heights ($z_h$) and radial length
 scales ($R_l$) using geometric description
 of the Galactic disk. 
 The distribution of warm
 dust in the Galactic disk is found to be characterised
 by $R_l < 6$ kpc and $60 \la z_h \la 100$ pc, 
 consistent with other studies.

 The location of the Galactic mid-plane as a
 function of longitude (in all 4 MSX bands),
 for stars as well as warm dust,
 has been searched for signatures of warp-like
 feature in their distribution,
 by fitting sinusoid with phase and amplitude as parameters. In every
 case, the warp signature has been detected. 
 Carrying out an identical analysis of the DIRBE/COBE data 
 (with lower angular resolution)
 in all its ten bands covering the entire infrared spectrum
 (1.25--240 $\mu$m), also leads to detection of
 warp signatures 
 with very  similar phase as found from the MSX data. 
 Our results have been compared with those from other
 studies.

}

\titlerunning{Warp of the Galactic disk as seen in mid infrared}
\authorrunning{S. Vig et al. }
\maketitle


\section{Introduction}
The study of structural features of the Galactic disk is useful in 
understanding the spatial distribution of its constituents, \textit{viz.} 
stars, interstellar gas and interstellar dust. Our Galaxy consists of the 
following kinematically distinct components: spheroidal (bulge, halo) and disk 
components (thin disk, thick disk).
The disk component of the Galaxy has 
been studied extensively using several infrared surveys.
 Kent et al. (\cite{Ke91}) 
obtained the three dimensional distribution of our Galaxy based on data  
at 2.4 $\mu$m from the Spacelab IRT. 
Freudenreich (\cite{Fr96}) used 
the Diffuse Infrared Background Experiment (DIRBE/COBE) data in
J, K, L and M bands to model the Galactic disk.

One of the subtler features of the Galaxy
is that of the warp structure, first detected through the 21 cm line 
emission of H I (Burke \cite{Bu57}; Kerr \cite{Ke57}). 
Various theories have been put forward to explain this phenomenon :
gravitational interaction with a companion galaxy
(Burke \cite{Bu57}; Kerr \cite{Ke57}; Hunter \& Toomre \cite{Hu69}; and others),
effect of massive halo (Binney \cite{Bi78}, \cite{Bi92}), 
misaligned halo-disc axes leading to cosmic infall 
(Ostriker \& Binney \cite{Os89}; Jiang \& Binney \cite{Ji99}),
intergalactic magnetic fields (Battaner et al. \cite{Ba90}, \cite{Ba91}),
dynamical friction between misaligned rotating halo and disc
(Debattista \& Sellwood \cite{De99}),
and more recently, intergalactic accretion flows onto the disc 
(Lopez-Corredoira et al. \cite{Lo02a}).
Porcel et al. (\cite{Po97}) have explained the Galactic stellar warp
using the tilted ring model of Rogstad et al. (\cite{Ro74}).
The warp in the stellar disk was detected 
from the distribution of sources in
the Infrared Astronomical Satellite (IRAS) 
Point Source Catalog (Djorgovski \& Sosin \cite{Dj89}). 
Using the DIRBE data, Freudenreich et al. (\cite{Fr94}), found 
evidence of warp for the stellar and the interstellar dust 
components of the Galaxy.
The old stellar populations near the Galactic plane have 
been analysed for Galactic flare and warp from the 
Two Micron All Sky Survey (2MASS) by Lopez-Corredoira et al. (\cite{Lo02b}).

Recently, the Midcourse Space Experiment (MSX) surveyed the entire 
Galactic plane in four mid infrared spectral bands (8--21 $\mu$m) with
high ($\sim$18\arcsec) angular resolution (Price et al. \cite{Pr01}). 
Taking advantage of the complete longitude coverage and 
superior angular resolution of this new dataset,
we have attempted to extract 
information about structure of the Galactic disk and search
for possible warp signature from the distribution of stellar as well as 
warm interstellar dust components of our Galaxy.
For a direct and meaningful comparison, an identical 
analysis has also been carried out with the DIRBE data,
albeit with poorer angular resolution. 

In Section 2, we discuss the MSX and DIRBE datasets used for the present study.
The data analysis procedures followed to
 extract structural features of the Galaxy and possible signatures of warp,
have been described along with the results in Section 3.
Based on a geometric description of the Galactic disk, 
a relation between the linear and the angular scale heights have 
been established in Section 4. 
Our results have been compared with others' in Section 5.
In Section 6, a brief summary has been presented.
 
\section{Data Sets}
The main objectives of the present study 
are to quantify parameters that characterise the structure 
of Galactic disk and search for possible signatures of warp 
from the distributions of stellar and warm 
interstellar dust components. 
The MSX data products have been used extensively.
In addition, DIRBE data have also been used for comparison.

\subsection{Midcourse Space Experiment (MSX)}
The MSX carried out mid infrared observations 
of the Galactic plane ($|$b$|\le 5^{\circ}$) in four bands A, C, D and E with 
$\lambda (\Delta\lambda)$ corresponding to 8.28 (3.36), 12.13 (1.72), 14.65 
(2.23), and 
21.34 (6.24) $\mu$m, respectively (Price et al. \cite{Pr01}). While the MSX Point 
Source Catalog, MSX PSC (Egan et al. \cite{Eg99}), has been used to study the 
distribution of point sources, the panoramic images in the four bands 
(pixel size 36\arcsec$\times$36\arcsec) have been used to study the diffuse emission. 
These images have already been corrected
for the zodiacal emission using an updated version of CBZODY model 
described in Kennealy et al. (\cite{Ke93}). The MSX PSC and the panoramic images 
were taken from IPAC ({\tt http://irsa.ipac.caltech.edu/applications/MSX/}) 
and the MSX team.

 We selected 4 samples of sources from the MSX PSC (one for each band),
 each based on the following criteria :
 (1) positive detection of the source in the particular band 
(quality flag $\geq 2 $); (2) source located within the latitude range 
 $|$b$|\leq2.75^{\circ}$, in order to avoid regions with non-uniform sky coverages; 
 and (3) sources brighter than the completeness limit of the catalogue
 for that band,
  \textit{viz.} 0.158 Jy, 1.58 Jy, 1.58 Jy 
and 3.16 Jy for 8, 12, 14 and 21 $\mu$m bands, respectively (as determined
from differential logN-logS distributions). 
The completeness limit in A band corresponds to 6.42 magnitude at 8 $\mu$m, 
 using the magnitude scale given by Egan et al. (\cite{Eg99}). 
Our samples consist of 164723, 25467, 18239 and 14254 sources in 8, 
12, 14 and 21 $\mu$m bands, respectively. The 8 $\mu$m band of MSX PSC
being the most sensitive, the corresponding sample is the largest. 
Using MSX PSC data, Lumsden et al. (\cite{Lu02}) have 
plotted known stars, planetary nebulae, H II regions
and young stellar objects 
(YSOs) on colour-colour diagrams in Figures 3, 4 \& 5 of
their paper. 
We have used a colour cut of the form 
\mbox{log$_{10}(F_{14}/F_{12})$ =  0.87 log$_{10}(F_{21}/F_{8})$ - 0.61}
(equation of a straight
line on a log-log plot; stars are located to the left
of this line) which separates the H II regions, etc from the stars,
as evident from their Figure 3 (right panel). Using this colour cut we
find the fraction of non-stellar objects to be $\sim 13\%$.
This is similar to IRAS Point Source Catalog, 
where $\sim 15 \%$ of the sources detected in both the
mid infrared bands (12 \& 25 $\mu$m) are 
non-stellar objects (Beichman et al. \cite{Be88}).

\subsection{Emission from DIRBE Sky Survey}
The DIRBE instrument onboard COBE spacecraft carried out
a full sky survey in 10 photometric bands centered 
at 1.25, 2.2, 3.5, 4.9, 12, 25, 60, 100, 140 and 240 $\mu$m, albeit with
limited angular resolution of $\sim 0.7^{\circ}$ (Boggess et al. \cite{Bo92}).
We have used the Zodi-subtracted Mission Average (ZSMA) maps here, which
were obtained from 
LAMBDA ({\tt http://lambda.gsfc.nasa.gov/product/cobe/dirbe\-\_products.cfm}).

\section{Analysis and Results}
\subsection{MSX Point Source Catalog (PSC) Sources}
\subsubsection{Estimated distribution of stellar types \& distance 
for our sample}

In order to estimate the distribution of stellar types 
of the mid infrared                  
sources in our sample and their distances, we have used 
a modified version of the Besan\c con model of stellar 
population synthesis (A. Robin, private communication). 
An earlier version of the model has been described in Robin et al. (\cite{Ro03}).
 The modified version of the model includes predictions of star counts 
in two mid infrared bands corresponding to the LW2 (7 $\mu$m) \& LW3 
(15 $\mu$m) filters of ISOCAM, which are crucial for the present study. 
This model 
synthesizes stars belonging to the populations of the thin disk, the thick 
disk, the stellar halo and the outer bulge including the warp and flare 
features.
This model has also been used to estimate the contribution of all stars,
including those not detected by MSX PSC, to the total mid IR emission
(see section 3.2.1).
Simulations have been carried out over 10$\times$10 deg$^2$ area for
$|$b$|<5^{\circ}$ and 30$^\circ< l \le 330^\circ$ on a grid with 30$^\circ$ spacing.
The simulations covered distances upto 20 kpc from the Sun and 
stars upto apparent magnitude of 14.0 at 7 $\mu$m.
The extinction due to interstellar dust has been neglected
in the model simulations.

As mentioned earlier, 
the completeness limit of MSX PSC for A band (8 $\mu$m)
corresponds to 6.4 magnitude at 8$\mu$m, [8].
This translates to a value of 6.5 magnitude at 7$\mu$m, [7], 
for the stellar types in our sample (the mean difference of the 
([7] - [8]) colour, is $\sim$ 0.1 mag; for various effective 
temperatures covered in our sample, as evident below).
Accordingly, we have considered the predicted star counts from the simulations
with [7] $<$ 6.5, to be representative of our MSX PSC sample.

The mean Galactocentric distance for our sample, is found to vary from $\sim$ 4 
kpc towards the Galactic centre ($l \sim$ 0$^{\circ}$) to $\sim$ 11 kpc towards the 
anti-centre direction ($l \sim$ 180$^{\circ}$). Most of the sources are dominated by 
K and M giants (31\% K \& 59\% M giants). The rest are AGB stars (6\%) and 
earlier than G dwarfs (4\%). 

\subsubsection{Latitude distribution of stellar sources and warp signature}
\begin {figure}
\begin {center}
\hskip -1cm
\includegraphics[height=4.5cm,width=4.5cm]{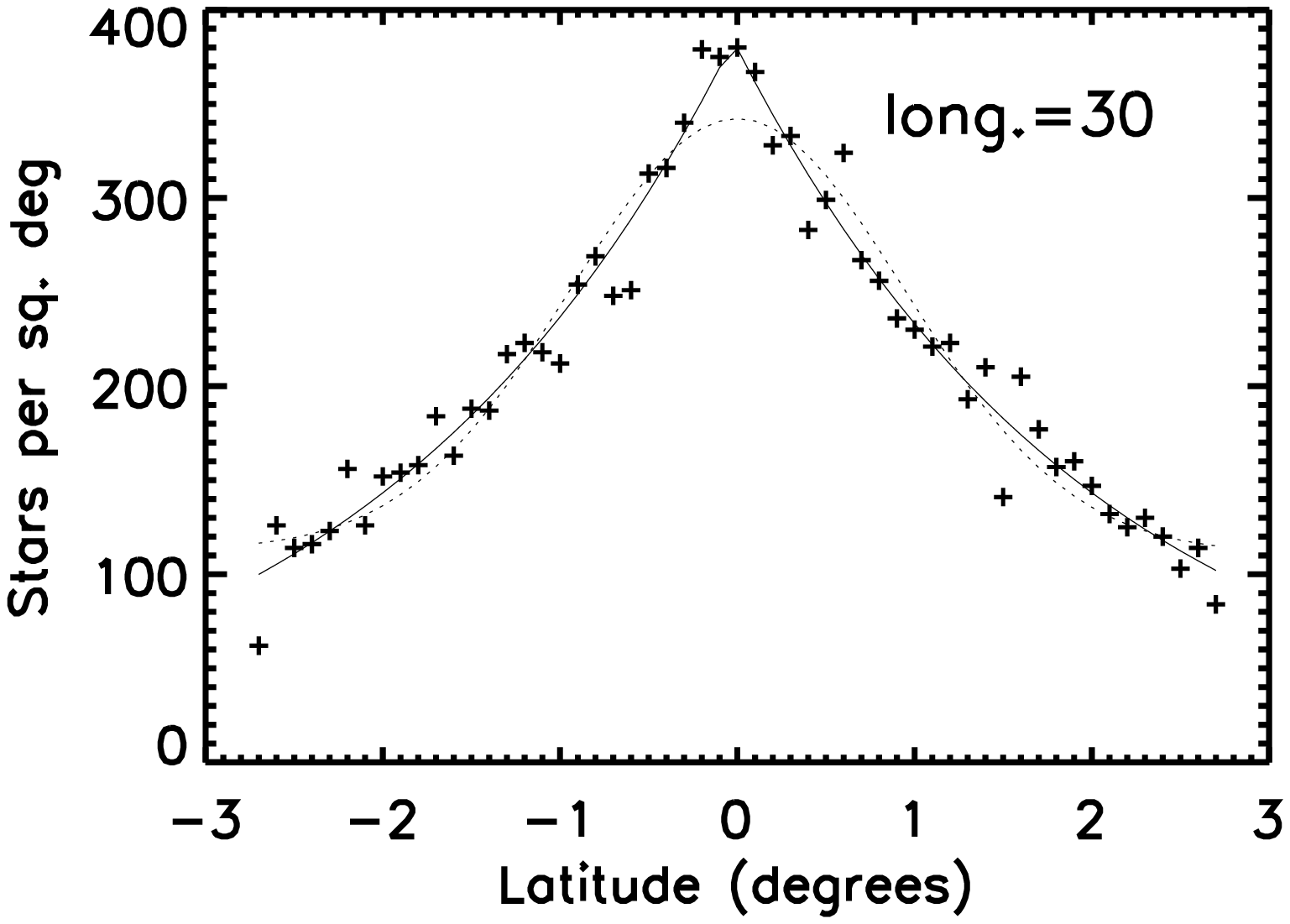}
\hskip -0.35cm
\includegraphics[height=4.5cm,width=4.5cm]{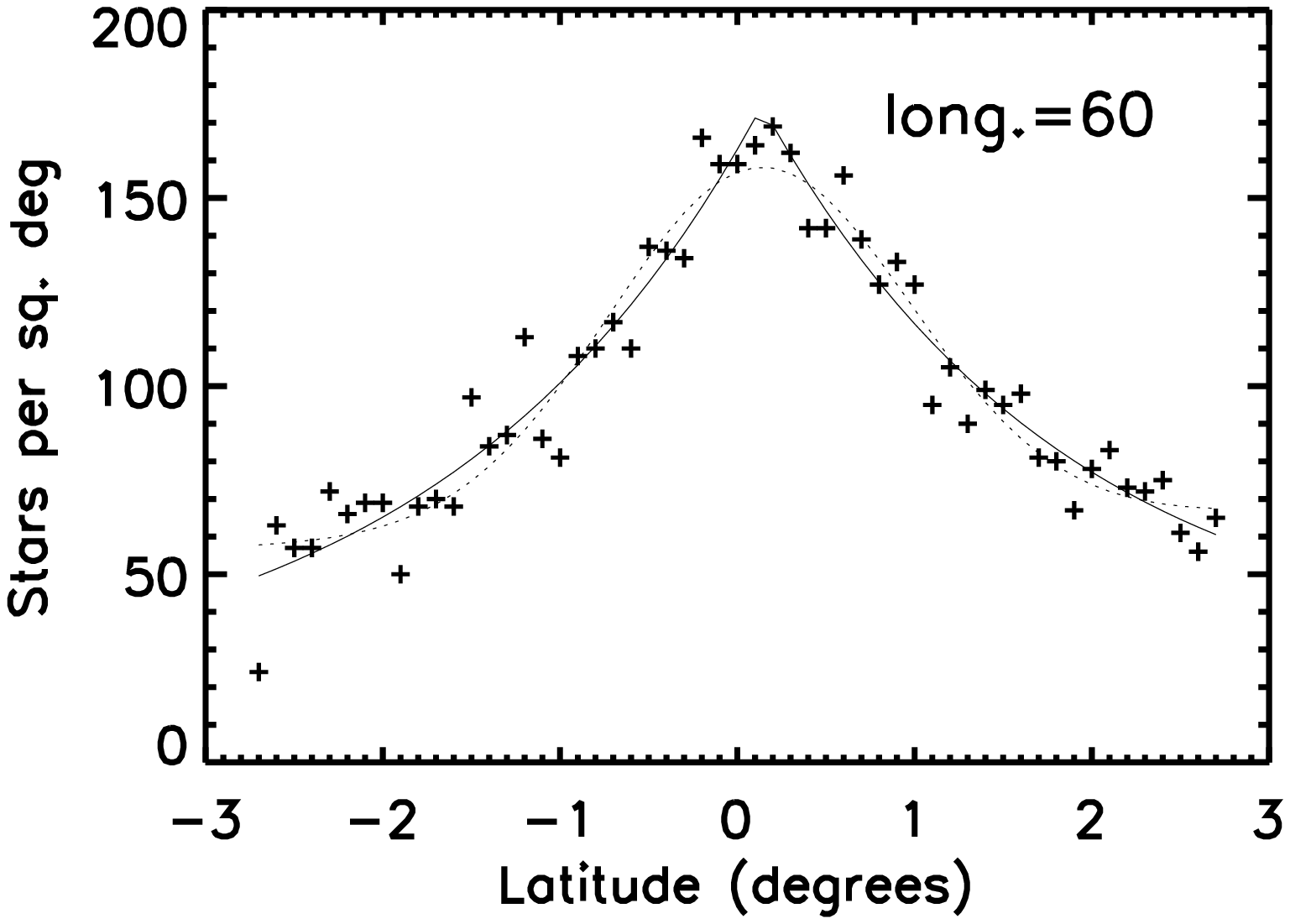}
\hskip -0.35cm
\includegraphics[height=4.5cm,width=4.5cm]{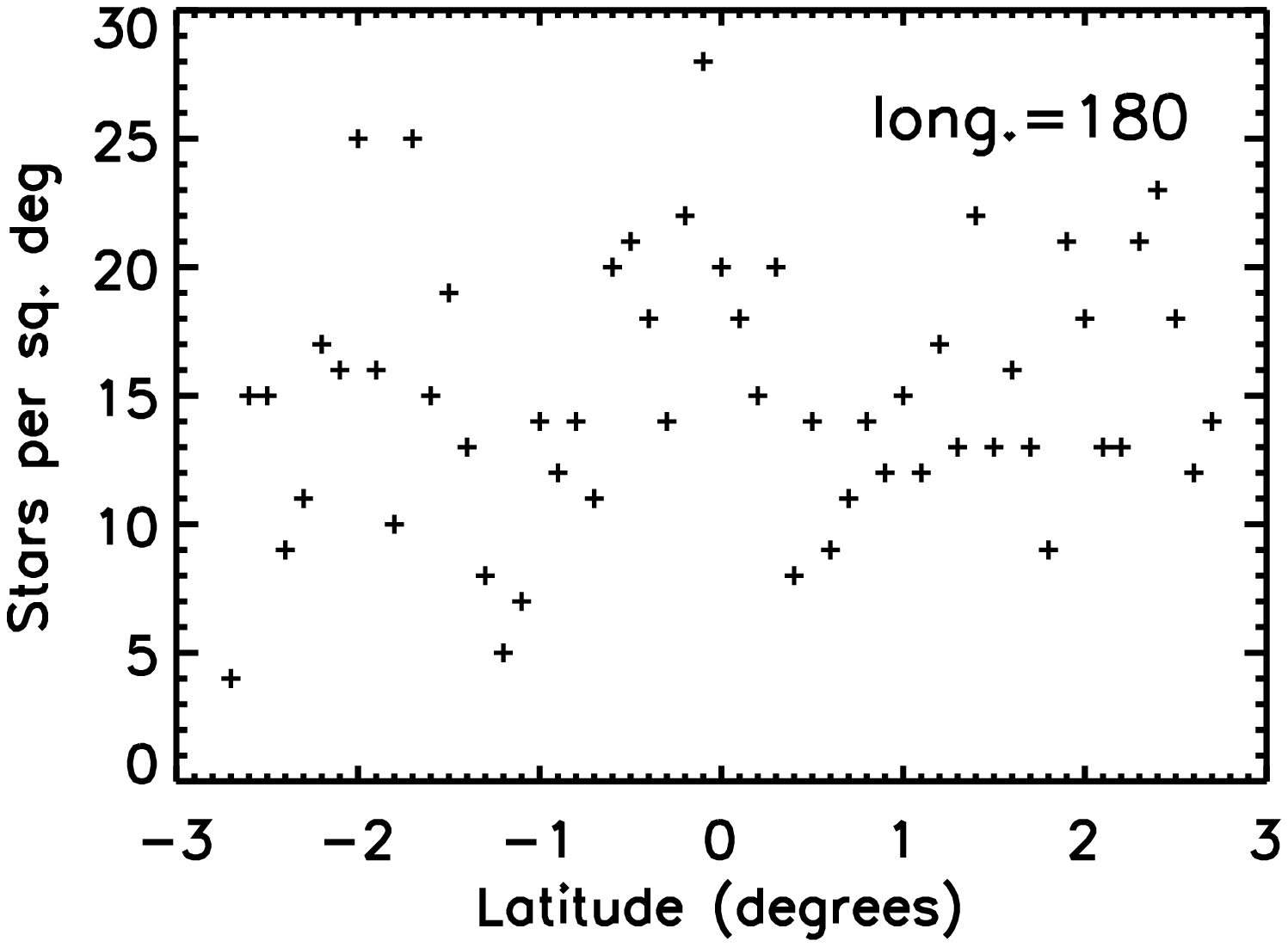}
\caption{The number density of stars as a function of latitude 
(latitude profile) detected in the A band of MSX PSC for the longitude bins 
 30$^{\circ}$, 60$^{\circ}$ and 180$^{\circ}$. 
The solid and the dotted lines represent best fit 
exponential and gaussian functions, respectively (for $l$= 30$^{\circ}$ \& 
60$^{\circ}$).
}
\vskip -0.6cm
\label{fig1}
\end{center}
\end {figure}

The structure of the Galactic disk has been studied 
 using the local distribution of stellar 
sources of our sample by constructing latitude profiles (LP) of stars for 36 
longitude bins, each 10$^\circ$ wide. The bin size along latitude 
is 0.1$^\circ$.  
The stellar density distribution in the Galaxy is 
generally represented by an exponential function of the form 
$n(z,R) \propto e^{-\frac{z}{z_h} - \frac{R}{R_l}}$, 
where $z$ is the height above the Galactic plane, 
$R$ is the Galactocentric distance, $z_h$ and $R_l$ represent the 
characteristic scale height and radial scale length, respectively. However, 
all relevant measurements are (approximately) heliocentric, 
and conveniently presented in Galactic coordinates ($l$, $b$).
Later (Section 4), we shall relate the linear and the angular
scales using a geometric description of the Galactic 
disk and show that exponential functions provide good fits to the 
latitude profiles too. 
Let $N(b)$ be the observed latitude profile representing
the number of stars as a function of latitude, $b$,
for a given longitude zone.
At first, we explore both the exponential and gaussian functions 
for fitting $N(b)$, as :
\begin{equation}           
 N(b) = \alpha_{e} e^{-\frac{|b-\beta_e|}{\gamma_e}} + 
\delta_e + \epsilon_e b 
\end{equation} 
\begin{equation}           
N(b) = \alpha_ge^{-(\frac{b-\beta_g}{\gamma_g})^2} + 
\delta_g + \epsilon_g b 
\end{equation}
where $\alpha$, $\beta$, $\gamma$, 
$\delta$ and $\epsilon$ are the free parameters.
 The subscripts $e$ and $g$ refer to the exponential and gaussian functions, 
respectively. 
The parameters $\alpha_e$ \& $\alpha_g$ represent the amplitude; 
$\beta_e$ \& $\beta_g$ represent the position of the Galactic 
midplane defined by the MSX PSC stars;
 $\gamma_e$ \& $\gamma_g$ represent the angular scale height;
$\delta_e$, $\delta_g$ and $\epsilon_e$, 
$\epsilon_g$ represent the background, if any, limited to linear 
dependence on latitude. Generally, $\beta = 0$ is expected if the peak 
of the latitude profile appears at $b$ = 0.
Any non-zero value signifies a shift. The fittings have been 
carried out for all the 36 longitude bins using a non-linear least squares 
method. For each of the four wavebands, 
good fits were obtained for the inner half
 of the Galactic plane towards the Galactic center \textit{i.e.} 
$-90^{\circ}<l\le90^{\circ}$ and very poor fits for the outer half, 
\textit{i.e.} towards 
the Galactic anti-center $90^{\circ} < l \le 270^{\circ}$. 
Predictions of the modified Besan\c con 
model also show the same trend for the distribution of stars 
([7] $<$ 6.5 mag). Figure ~\ref{fig1} shows the latitude
profiles (MSX PSC, A band) for three representative 
longitude bins (30$^{\circ}$, 60$^{\circ}$ \& 180$^{\circ}$), 
including functional fits for the
two longitudes in the inner Galaxy.
For all the four bands, the exponential function fits the latitude 
profiles better than the gaussian function, which is
consistent with our understanding of the Galactic disk. 
Consequently, the results pertaining to the exponential function
only have been presented here.
Table ~\ref{tab1} lists the mean value and standard deviation of the parameter $\gamma$
for each waveband, as obtained from
the 18 longitude bins covering the inner Galactic plane.


\begin{table}
\begin{center}
\caption{Characteristic angular scale, $\gamma$, of latitude profiles
obtained by fitting exponentials to the MSX PSC sample, MSX emission and 
DIRBE emission for the 18 longitude bins covering 
$-90^{\circ} < l \le 90^{\circ}$.}
\label{tab1}
\begin{tabular}{|c c c|} \hline\hline
 Wavelength  &  \multicolumn{2}{c|}{Exponential} \\ 
 $\mu$m (Band) &  \multicolumn{2}{c|}{$\gamma_e$ in $\deg$(s.d.)} \\ \hline
 & MSX PSC & MSX Diffuse emission\\ \hline
 8.3  (A)  & $1.29 (0.67)$ &  $1.69 (0.45)$ \\
 12.1 (C)  & $1.27 (0.57)$ &  $0.92 (0.41)$ \\
 14.7 (D)  & $1.25 (0.70)$ &  $0.71 (0.22)$ \\
 21.3 (E)  & $1.04 (0.96)$ &  $0.56 (0.28)$ \\ \hline
 & & DIRBE Emission \\ \hline
 1.25 && $1.94 (1.18)$  \\
 2.2  && $2.74 (0.78)$  \\
 3.5  && $1.93 (0.54)$  \\
 4.9  && $1.57 (0.47)$  \\
 12   && $1.29 (0.67)$  \\
 25   && $1.00 (0.46)$  \\
 60   && $0.92 (0.33)$  \\
 100  && $1.23 (0.77)$  \\
 140  && $1.19 (0.69)$  \\
 240  && $1.30 (0.87)$  \\ \hline \hline
\end{tabular}
\end{center}
\end{table}

\begin {figure}
\begin {center}
\includegraphics[height=9.0cm]{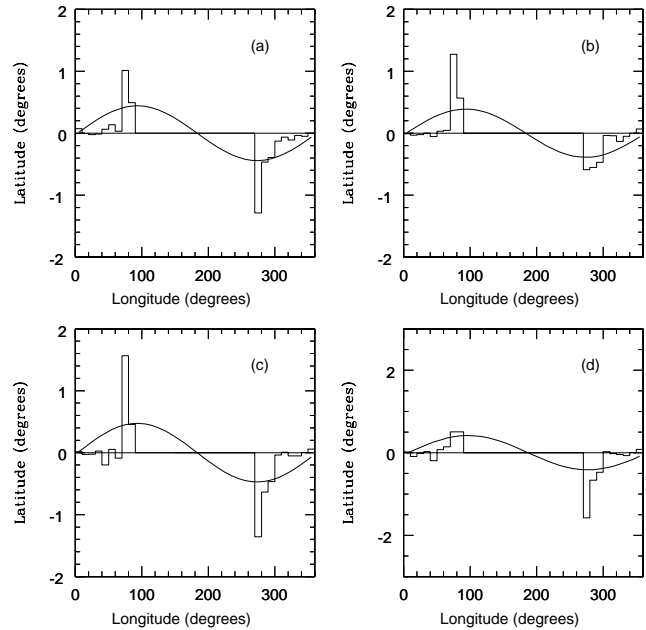}
\caption{The variation of the 
location of Galactic midplane with longitude, $\beta(l)$, as 
obtained from fitting exponential   
to the latitude profiles (LP) for our MSX PSC sample (histogram)
for the inner Galaxy in 4 MSX bands. 
The solid lines refer to the extracted WS
in (a) 8 $\mu$m, (b) 12 $\mu$m, (c) 14 $\mu$m, and (d) 21 $\mu$m bands.
}
\vskip -0.6cm
\label{fig2}
\end{center}
\end {figure}

Next we explore possible signatures of warping in the Galactic disk.
To quantify any evidence for the Galactic warp 
from the local distribution of 
stars, we search for systematic variation of $\beta$s with longitude. 
Djorgovski \& Sosin (\cite{Dj89}), used a sinusoid function to fit the 
distribution of IRAS point sources. A similar approach 
has been adopted here, but without the additive constant term, thereby
assuming that the Sun is located at the local midplane.
Thus, the function used to fit the longitude variation of the 
profile peak is :
\begin{equation}
\beta(l) = \beta_o \times sin(l-\phi)   
\end{equation}
where $\beta$ represents the best fit to  
$\beta_e(l)$ of eqn (1) obtained for $-90^{\circ}<l\le90^{\circ}$. 
Although useful, this is admittedly an approximate approach
due to the fact that the measurements are heliocentric
(hence, lines of sight along different longitudes
cover different galactocentric distances of the disc), 
whereas a true sinusoid due to any warp feature is expected 
only for a galactocentric observer. 
We crudely relate the
parameters 
$\beta_{\circ}$ and $\phi$ to quantify the local 
warp signature (hereafter WS) indicating amplitude and phase 
respectively. This phase, $\phi$, corresponds to the position 
angle (longitude) of the line of intersection between the $b$=0 plane and the 
warped Galactic plane. In order to carry out a quantitative comparison 
with a `null' hypothesis (i.e. WS absent), we also fit a 
constant function,  
$\beta(l) = k $,  
to the same data sets. 
The resulting best fit parameters along with their standard errors
have been presented in Table ~\ref{tab2}. 
The data and the fitted functions are displayed in Figure ~\ref{fig2}. The WS 
has been clearly detected in all the 4 bands as is evident from the ratio of 
$\chi^2$s per degree of freedom ($R > 1$).
On repeating our analysis with a larger sample of MSX PSC sources
with latitude cut relaxed to $|b|<$ 5$^{\circ}$,
again we detect WS with very similar parameters.

Since only the inner half of the Galactic plane could be 
studied using the above approach, 
we pursue additional measures of statistical nature 
to quantify the location of the Galactic midplane, 
which could be effective for the entire longitude range.
We have explored two statistical measures of central tendency, viz.,
the mean and the median latitude. 
The mean refers to the average latitude of the 
MSX PSC sources of our sample for a particular longitude bin.
 The median latitude refers to a value such 
that the total number of stars located at higher and lower 
latitudes are equal. The mean is a good measure of central tendency for 
roughly symmetric distributions but can be misleading for
skewed distributions. The median should be more 
informative for such distributions.

\begin {figure}
\begin {center}
\includegraphics[height=9.0cm]{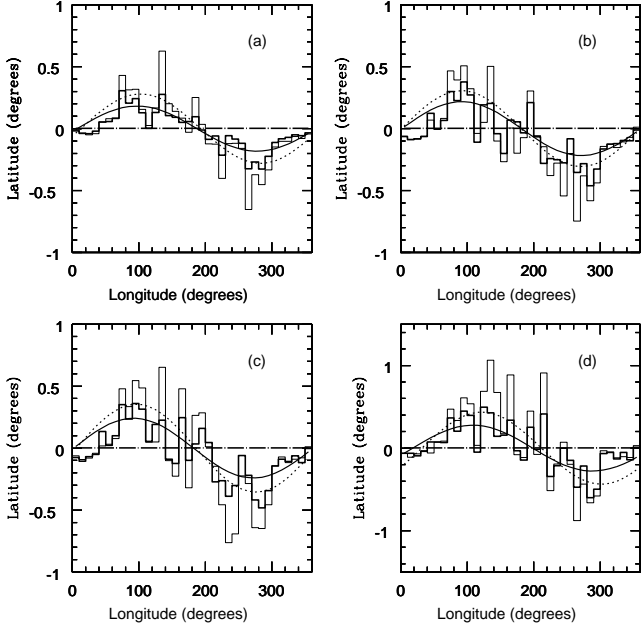}
\caption{
The variation of the SIs of the Galactic midplane, \textit{viz.}
 mean (thick line histogram) and median (thin line histogram) 
latitude as a function of longitude for our MSX PSC sample
in the 4 MSX bands :
(a) 8 $\mu$m, (b) 12 $\mu$m, (c) 14 $\mu$m, and (d) 21 $\mu$m.
The solid and the dotted curves represent the extracted WS 
from the mean and median latitudes, respectively.}
\vskip -0.6cm
\label{fig3}
\end{center}
\end {figure}

The mean and median latitudes for different
Galactic longitude bins have been displayed as histograms in Figure ~\ref{fig3}. 
A systematic deviation from the nominal Galactic plane ($b$ = 0)
is clearly visible in this figure. 
Both the statistical indicators (SI) are positive over one 
half of the longitude range $\sim 0^{\circ}-190^{\circ}$ and 
negative in the other 
half. The positive and the negative peaks occur at  
$l \sim 90^{\circ}-100^{\circ}$
 and $\sim260^{\circ}-280^{\circ}$, respectively. This is 
the general trend for all the 4 bands of MSX. Similar to the case 
of $\beta$ above for the exponential fits (eqn. 3), we fit a sinusoid
(with $b_{\circ}$ and $\phi$ as parameters) to the 
mean and the median latitudes giving equal weightage to each longitude bin. 
Once again, the `null' hypothesis is also explored for comparison. 
The resulting best fit parameters along with standard errors, 
for all the cases have been presented in Table ~\ref{tab3}. 
The corresponding fitted functions for respective bands
 have been overplotted in Figure ~\ref{fig3}.
The signature of warp is clearly indicated 
from both the statistical indicators (SIs) for all the 4 MSX bands. 
 This is also evident from the ratio of the  $\chi^{2}$s ($R > 1$). 
These findings are
statistically further strengthened by the fact that the values of the phase, 
$\phi$, are consistent (within errors) among all four bands of MSX and also 
matches that expected from the known features of the Galactic stellar warp 
(Djorgovski \& Sosin \cite{Dj89}; see Section 5.2.1). 
The SI approach seems to be more successful than LP. This could be because the 
statistical determinations of central tendencies are inherently more
immune to fluctuations than procedures involving fitting of a 
specific function.

The amplitude of the sinusoid is lower for the SI 
than the $\beta$s obtained from the fitting of the latitude profile. 
Among the mean and median, the displacement is larger for the latter case than 
the former. 
The mean provides a better quality fit to sinusoid than the median,
based on comparison of corresponding reduced $\chi^2$s.

\subsection{Diffuse emission from MSX images}
Diffuse mid infrared emission is expected from the warmer (T $\ga$ 100 K) 
interstellar dust heated by nearby luminous stars. The emission in the 
mid infrared (obtained from the MSX panoramic images) contains 
contributions from stellar sources also. In the next subsection, we estimate 
the contribution of point sources, which turns out to be
a negligible fraction of the total emission detected.

\subsubsection{Contribution of point sources to the MSX data}
The panoramic images from MSX
includes the contribution from the point sources, some of which have been
detected and are catalogued in the MSX PSC. 
The sources from our MSX PSC sample
contribute nearly 4.5\%, 1.9\%, 3.5\% and 2.9\% of the total mid 
infrared emission in the 8, 12, 14 and 21 $\mu$m bands, respectively.
(The corresponding fractions considering $all$ the MSX PSC sources
are : 4.7\%, 1.9\%, 3.8\% \& 3.1\%.) 
However, these correspond only to the sources brighter than the
 completeness limit of the catalog. In other words, there are (1)
unresolved sources (MSX angular resolution is $\sim$18\arcsec) and (2) sources
fainter than
the sensitivity limit, contributing to the intensity in these images, apart
from the cataloged point sources. 
To estimate the contribution of unresolved and faint stellar sources in the
mid infrared MSX data, we have used 
the modified Besan\c con model described earlier (Section 3.1.1). 
The contributions from stars in the selected latitude range 
for various longitude zones have been determined from the model simulations. 
Integrated fluxes at 7 $\mu$m from all stars within a heliocentric distance 
of 20 kpc and apparent brightness higher than m$_{lim}$ ([7] $<$ m$_{lim}$,
where m$_{lim}$ = 6.5, 7.0, ...., 14.0), 
have been determined.
It was noted that beyond m$_{lim}$=12, the 
increase in the value of the integrated flux was insignificant ($\sim 2\%$). 
 The ratio of integrated fluxes for sources with [7]$<14.0$ (representing 
the entire Galaxy) and [7]$<6.5$ 
(MSX PSC completeness at A band; section 3.1.1) is found to be $\sim$ 1.4.
Carrying out similar model simulations at 
15 $\mu$m and considering the completeness of MSX PSC at D band,
we find the ratio of corresponding integrated fluxes is 
$\sim 1.3$. It may be noted that the values of these
ratios are in fact upper limits, since the
interstellar extinction has been neglected in the model simulations. 
The above implies that, the contribution of $all$ stars towards 
the total emission is $\sim 6.3\%$ \& $\sim4.6$\% 
 for 8 \& 14 $\mu$m MSX bands respectively.
If we adopt a value of 1.4 for the ratio of integrated fluxes for the 
12 \& 21 $\mu$m bands also, then the stellar contributions 
in the respective bands would be $\sim2.7\%$  \& $\sim4.1\%$.
Therefore, we conclude that most of the total 
mid infrared emission detected in the MSX bands are of diffuse origin. 
From a study of the diffuse infrared emission
(for $|$b$|>10^{\circ}$) of our Galaxy using the IRAS data, 
Boulanger \& P\'erault (\cite{Bo88}) found that the point sources 
account for 10\% $\pm$ 2\% of the 12 $\mu$m flux.

\subsubsection{Latitude distribution of diffuse emission and warp 
signature from MSX data}
\begin {figure}
\begin {center}
\hskip -1cm
\includegraphics[height=5.0cm,width=5.0cm]{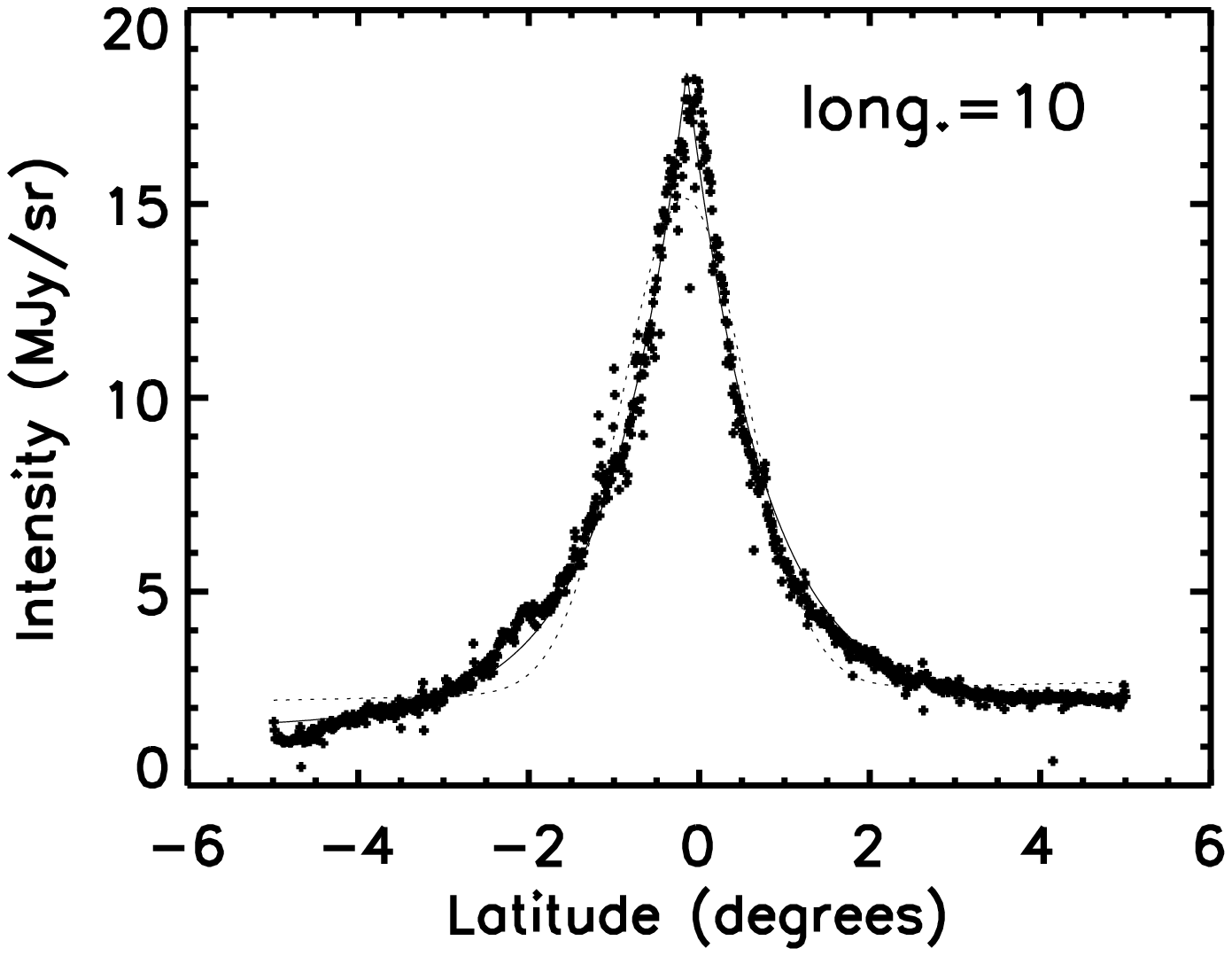}
\hskip -0.35cm
\includegraphics[height=5.0cm,width=5.0cm]{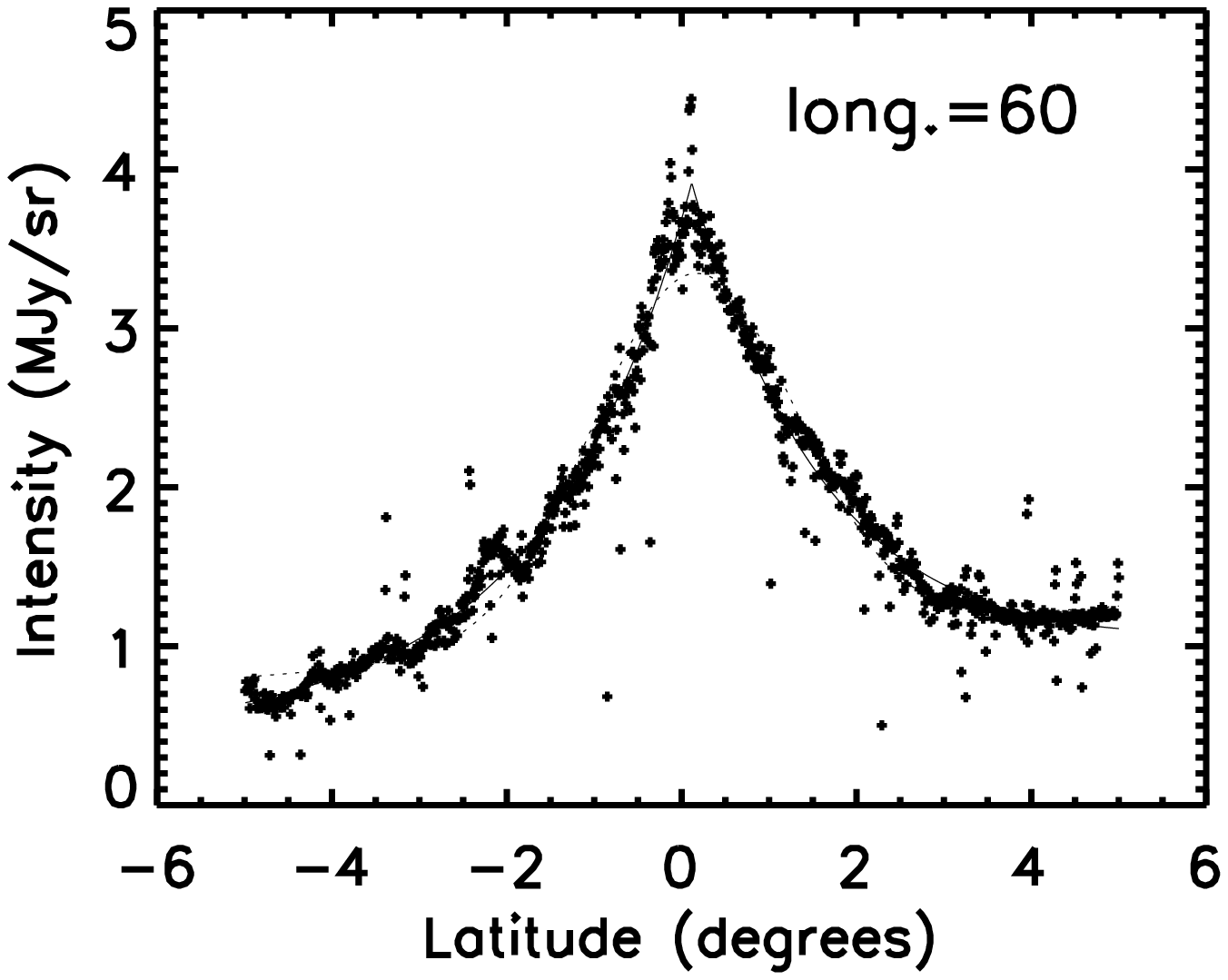}
\hskip -0.35cm
\includegraphics[height=5.0cm,width=5.0cm]{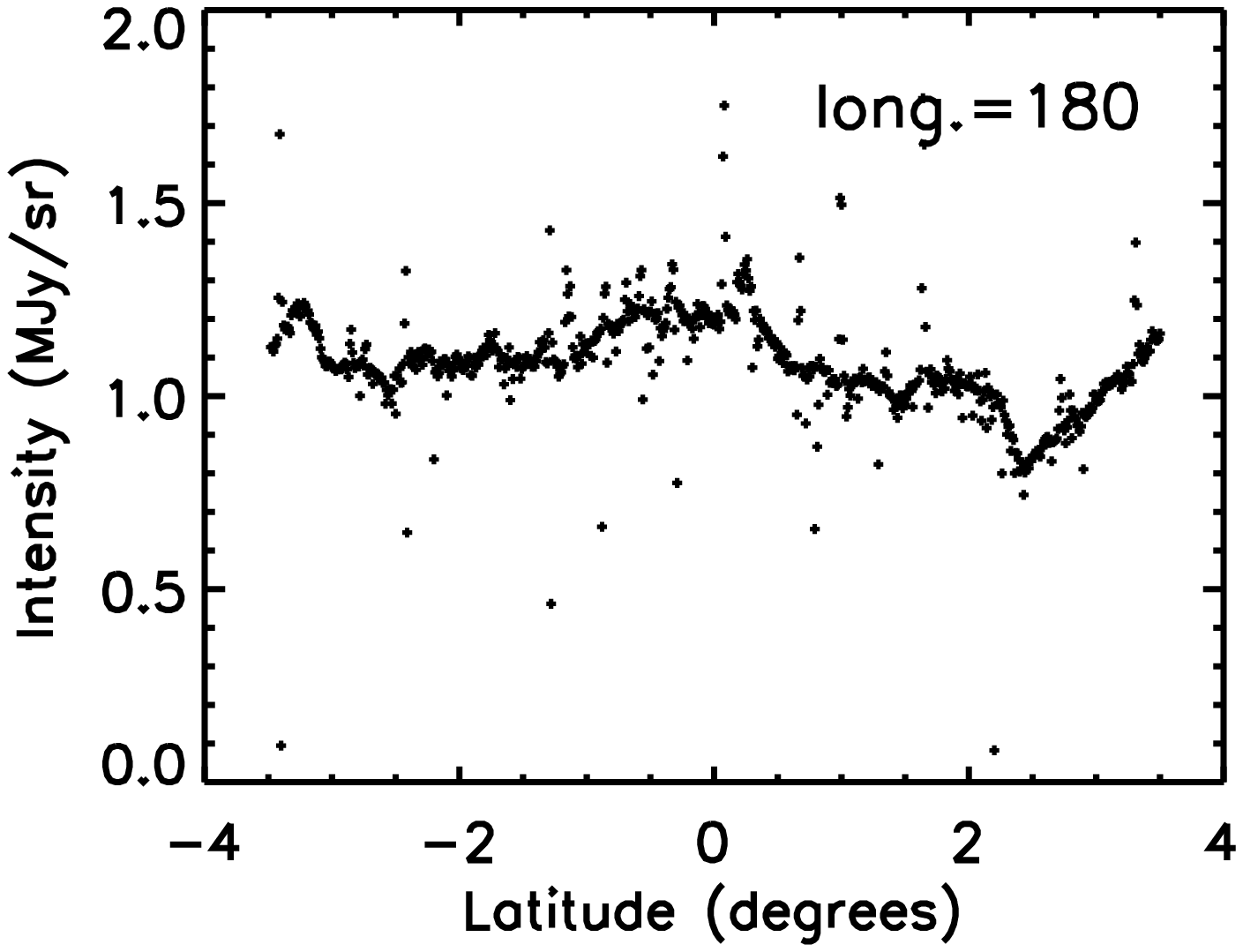}
\caption{Latitude profiles of MSX A band intensity for longitude bins
10$^{\circ}$, 60$^{\circ}$ and 180$^{\circ}$.
The solid and the dotted lines represent best fitting exponential and gaussian
functions, respectively (for $l$=10$^{\circ}$ \& 60$^{\circ}$).}
\vskip -0.6cm
\label{fig4}
\end{center}
\end {figure}

The latitude distribution of diffuse emission in the 4 MSX bands, 
has been studied following an analysis similar  
to that used for the stars (Section 3.1.2), but for 
$|$b$|\le5^{\circ}$. Exponential and gaussian functions (see Eqns (1) and (2)) 
have been fitted 
to the latitude profiles of intensity at respective bands. 
The fits are found to be 
good for the inner Galaxy ($-90^{\circ}<l\le90^{\circ}$) 
but very poor towards the outer Galaxy, similar to the stellar case. 
The latitude profiles of intensity in A band for three representative 
longitude bins (10$^{\circ}$, 60$^{\circ}$ \& 180$^{\circ}$)
have been presented in Figure ~\ref{fig4}, 
which also displays the fitted functions for the
two bins in the inner Galaxy.
 Again, we find that the exponential provides better fits than 
the gaussian for all the cases, as expected. 
Henceforth, we discuss the exponential only. The average values 
(\& standard deviation) of 
the angular scale height parameter, $\gamma_{e}$, 
 from the 18 longitude bins, have been listed in Table ~\ref{tab1}.
It may be commented here that, the inferred $\gamma_{e}$ should be
immune to any residuals from the zodiacal subtraction, because :
(i) the angular scale of zody emission is much larger than typical
$\gamma_{e}$; \& (ii) the background terms (constant \& linear in $b$)
should absorb all components varying slowly with $b$.  

\begin {figure}
\begin {center}
\includegraphics[height=9.0cm]{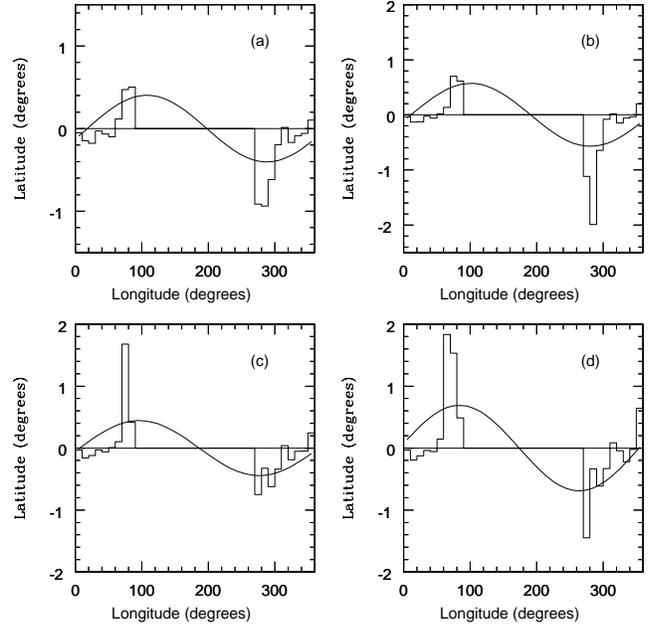}
\caption{The variation of the location of Galactic midplane with longitude as
obtained by fitting exponential to the latitude profile (LP) of 
the total (diffuse) emission measured in 4 MSX bands (histogram)
for the inner Galaxy. 
The solid lines refer to the extracted WS in
(a) 8 $\mu$m, (b) 12 $\mu$m, (c) 14 $\mu$m, and (d) 21 $\mu$m bands.
}
\vskip -0.6cm
\label{fig5}
\end{center}
\end {figure}

In order to search for WS in the diffuse emission, 
we fitted  sinusoids to the locations of
profile peak at various longitudes, $\beta^{diff}_{e}(l)$, for all
4 MSX bands (similar to the stellar case; Eqn. 3). 
Once again, the WS has been detected.
The best fit values of the parameters of the WS, viz., amplitude and phase 
are listed in Table ~\ref{tab2}. The $\beta^{diff}_{e}(l)$s and corresponding 
best fitting sinusoids are presented in Figure ~\ref{fig5}.

\begin {figure}
\begin {center}
\includegraphics[height=9.0cm]{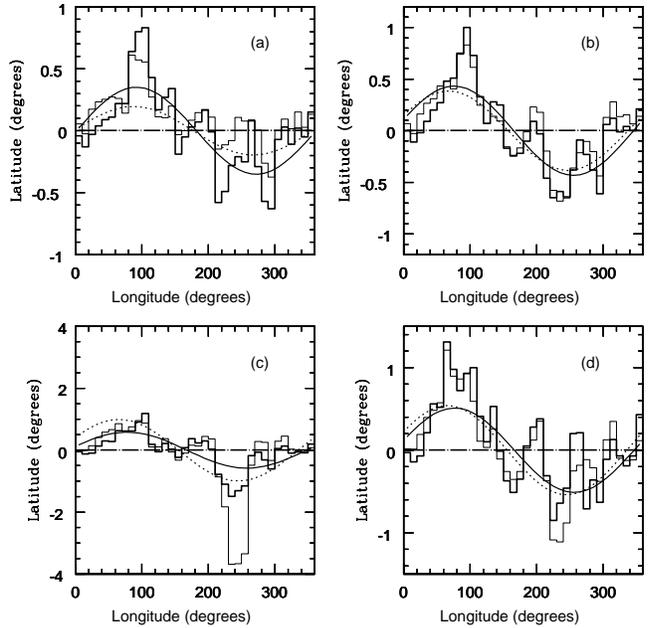}
\caption{The variation of the SIs of the Galactic midplane, 
\textit{viz.} intensity weighted mean ($B_{intwt}$; thick line histogram) and 
median ($B_{med}$; thin line histogram) latitudes as a function of longitude 
for the diffuse emission in the 4 MSX bands :
 (a) 8 $\mu$m, (b) 12 $\mu$m,
(c) 14 $\mu$m, and (d) 21 $\mu$m. The solid and the dotted curves
refer to 
the extracted WSs for the  
intensity weighted and median latitudes, respectively.
 }
\vskip -0.6cm
\label{fig6}
\end{center}
\end {figure}

We have also pursued the statistical approach (SI; similar to the stellar case)
to extend the search for WS over the entire Galactic longitude range. 
Two measures of central tendencies have been defined for the 
latitude distribution of intensity at various $l$. 
The intensity weighted central latitude
$B_{int wt}(l)$ of the diffuse emission is defined by :
\begin{equation}           
B_{int wt}(l)=\sum_{b_i=-5^{\circ}}^{5^{\circ}} b_if_i /\sum_{b_i=-5^{\circ}}^{5^{\circ}} f_i
\end{equation}
where $b_i$ and $f_i$ are the latitude and the diffuse emission intensity for 
the pixel $i$, considering all pixels within the particular longitude bin $l$. 
The median latitude, $B_{med}(l)$, is defined to be
that latitude, on both sides of which the areas under the intensity vs
latitude curve are equal,
\begin{equation}            
\sum_{b_i=-5^{\circ}}^{B_{med}(l)}f_i\Delta b =
\sum_{b_j>B_{med}(l)}^{5^{\circ}}f_j\Delta b
\end{equation}
where $\Delta b$ is the pixel size
along latitude (36\arcsec). This equation (5) is
 solved to determine $B_{med}$ for each of the 36 longitude bins. 
The statistical indicators (SI) for the Galactic midplane, $B_{intwt}$ \& $B_{med}$, 
 as a function of Galactic longitude for each MSX band are presented
in Figure ~\ref{fig6}. These SIs for diffuse emission have also been fitted with sinusoids 
to search for WS.


\begin{table}
\begin{center}
\caption{The values of the best fit parameters of the 
warp signature, WS (sinusoid 
with amplitude $\beta_{\circ}$ \& phase $\phi$) to the 
longitude variation (-90$^{\circ}<l\le90^{\circ}$) of 
locations of the 
peaks of latitude profile, LP, obtained from the exponential fitting to :
 the MSX PSC sources ($|$b$|\leq 2.75^{\circ}$), 
diffuse emission in MSX bands ($|$b$|\leq 5^{\circ}$), and 
emission in DIRBE bands ($|$b$|\leq 5^{\circ}$). 
The free parameter, $k$, for the null hypothesis is also tabulated. 
The ratio of $\chi^2$ per degree of freedom,
R, for the best fitting null case and
the WS case, is also presented. A value of R$>1$ indicates
that the sinusoid is a better fit.
}
\label{tab2}
\begin{tabular}{|c c c c c|} \hline\hline
 $\lambda$ & $\beta_{\circ}$ & $\phi$ & $k$ & R  \\ 
($\mu$m)  & (deg) & (deg) & (deg) & \\ \hline
\multicolumn{5}{|c|}{MSX PSC} \\ \hline
 8.3  &  $0.44\pm0.11$ & $3.1\pm14$ &  $-0.04\pm0.10$ &$1.95$ \\
 12.1 & $0.39\pm0.10$ & $3.5\pm16$ & $-0.00\pm0.10$ & $1.80$ \\
 14.7 & $0.47\pm0.15$ & $3.2\pm19$ &$-0.04\pm0.13$ & $1.53$ \\
 21.3 & $0.42\pm0.13$ & $6.7\pm18$ &$-0.10\pm0.11$ & $1.53$ \\ \hline
\multicolumn{5}{|c|}{MSX Diffuse Emission} \\ \hline
 8.3  & $0.40\pm0.10$ & $18\pm14$  & $-0.13\pm0.09$ &$1.81$ \\
 12.1 & $0.57\pm0.16$ & $11\pm17$  & $-0.15\pm0.14$ &$1.58$ \\
 14.7 & $0.44\pm0.14$ & $6.2\pm18$ & $-0.13\pm0.12$ & $1.60$ \\
 21.3 & $0.69\pm0.19$ & $-6.3\pm16$ & $0.07\pm0.17$  & $1.75$ \\ \hline
\multicolumn{5}{|c|}{DIRBE Emission} \\ \hline
 1.25 &  $1.33\pm0.37$ & $72\pm16$ &$-0.71\pm0.29$ & $1.30$ \\
 2.2 & $0.51\pm0.08$ & $19\pm9.5$ & $-0.08\pm0.10$ &$3.11$ \\
 3.5 & $0.38\pm0.06$ & $15\pm9.4$ & $-0.04\pm0.07$ &$3.20$ \\
 4.9 & $0.34\pm0.07$ & $15\pm11$ & $-0.04\pm0.07$ &$2.63$ \\
 12  & $0.36\pm0.11$ & $18\pm17$ & $0.12\pm0.09$ &$1.50$ \\
 25  & $0.39\pm0.16$ & $11\pm23$ & $-0.06\pm0.12$ &$1.30$ \\
 60  & $0.45\pm0.15$ & $12\pm19$ & $-0.07\pm0.12$ &$1.46$ \\
 100 & $0.40\pm0.11$ & $13\pm16$ & $-0.10\pm0.10$ &$1.65$ \\
 140 & $0.37\pm0.11$ & $16\pm17$ & $-0.12\pm0.09$ &$1.52$ \\
 240 & $0.37\pm0.11$ & $16\pm17$ & $-0.12\pm0.09$ &$1.48$ \\
 \hline\hline
\end{tabular}
\end{center}
\end{table}



\begin{table}
\begin{center}
\caption{The values of the best fit parameters of the WS (sinusoid 
with amplitude $b_{\circ}$ \& phase $\phi$) 
to the longitude variation (0$^{\circ}<l\le 360^{\circ}$) of the
statistical indicators SI of 
Galactic midplane, viz., the mean \& median for :
 MSX PSC stars ($|$b$|\leq 2.75^{\circ}$),
 intensity 
weighted \& median latitude of emission for MSX and 
 DIRBE bands ($|$b$|\leq 5^{\circ}$). 
The free parameter, $k$, for the null hypothesis is 
also tabulated. The ratio of $\chi^2$ per degree of freedom, R,
 for the best 
fitting null case and the WS case, is 
also presented 
(R$>1$ indicates the sinusoid to be a better fit).
} 
\label{tab3}
\begin{tabular}{|c c c c c|} \hline\hline
 $\lambda$ &  $b_{\circ}$ & $\phi$ & $k$ & R \\ 
 $\mu$m  &  (deg) & (deg) & (deg) &  \\ \hline
 \multicolumn{5}{|c|}{MSX PSC : Mean, Median} \\ \hline
 8.3  & $0.18\pm0.02$ & $7.3\pm5.9$ & $-0.01\pm0.02$ &$3.75$ \\
      & $0.28\pm0.03$ & $13\pm7.3$ & $-0.02\pm0.04$ &$2.78$ \\
 12.1 & $0.22\pm0.03$ & $1.3\pm8.5$ & $-0.05\pm0.03$ &$2.15$ \\
      & $0.31\pm0.05$ & $1.1\pm9.2$ & $-0.07\pm0.05$ &$1.98$ \\
 14.7 & $0.24\pm0.03$ & $1.9\pm7.8$ & $-0.04\pm0.04$ &$2.46$ \\
      & $0.35\pm0.06$ & $6.0\pm9.7$ & $-0.05\pm0.06$ &$1.95$ \\
 21.3 & $0.28\pm0.04$ & $18\pm9.0$ & $0.01\pm0.04$ &$2.16$ \\
      & $0.43\pm0.08$ & $32\pm10$ & $0.07\pm0.07$ &$1.83$
\\ \hline
 \multicolumn{5}{|c|}{MSX Diffuse emission : Intensity weighted, Median} \\ \hline
 8.3  & $0.35\pm0.05$ & $1.8\pm8.9$    & $0.02\pm0.06$  &$2.19$ \\
      & $0.19\pm0.04$ & $-2.7\pm13$  & $0.11\pm0.03$  &$1.18$ \\
 12.1 & $0.43\pm0.06$ & $-15\pm7.6$ & $0.01\pm0.06$  &$2.64$ \\
      & $0.38\pm0.06$ & $-23\pm8.3$  & $0.06\pm0.06$  &$2.30$ \\
 14.7 & $0.58\pm0.10$ & $-12\pm9.6$ & $-0.03\pm0.10$ &$2.01$ \\
      & $0.99\pm0.22$ & $-25\pm13$ & $-0.16\pm0.19$ &$1.51$ \\
 21.3 & $0.51\pm0.09$ & $-14\pm10$ & $0.08\pm0.08$ &$1.82$ \\
      & $0.54\pm0.08$ & $-25\pm8.7$  & $0.05\pm0.08$  &$2.22$ \\ \hline
 \multicolumn{5}{|c|}{DIRBE emission : Intensity weighted, Median} \\ \hline
 1.25 & $0.22\pm0.07$ & $68\pm17$ & $-0.08\pm0.05$ &$1.18$ \\
      & $0.39\pm0.11$ & $65\pm17$ & $-0.24\pm0.08$ &$1.05$ \\
 2.2  & $0.20\pm0.05$ & $34\pm14$ & $-0.04\pm0.04$ &$1.41$ \\
      & $0.30\pm0.08$ & $30\pm15$ & $-0.14\pm0.06$ &$1.20$ \\
 3.5  & $0.26\pm0.04$ & $15\pm9.8$ & $ -0.03\pm0.04$ &$1.95$ \\
      & $0.37\pm0.07$ & $11\pm11$ & $-0.14\pm0.06$ &$1.55$ \\
 4.9  & $0.30\pm0.04$ & $6.8\pm8.6$ & $-0.03\pm0.05$ &$2.27$ \\
      & $0.41\pm0.07$ & $3.8\pm9.3$ & $-0.14\pm0.06$ &$1.80$\\
 12   & $0.39\pm0.05$ & $-13\pm7.0$ & $-0.04\pm0.06$ &$2.89$ \\
      & $0.51\pm0.08$ & $-11\pm8.9$ & $-0.19\pm0.08$ &$1.85$\\
 25   & $0.35\pm0.04$ & $-6.8\pm7.3$ & $-0.01\pm0.05$ &$2.79$ \\
      & $0.49\pm0.08$ & $-3.8\pm9.3$ & $-0.16\pm0.07$ &$1.86$ \\
 60   & $0.54\pm0.07$ & $-3.8\pm7.1$ & $-0.03\pm0.08$ &$2.89$ \\
      & $0.68\pm0.11$ & $-6.1\pm9.6$ & $0.18\pm0.11$ &$1.87$\\
 100  & $0.49\pm0.06$ & $-8.2\pm7.0$ & $-0.04\pm0.07$ &$2.89$ \\
      & $0.64\pm0.09$ & $-9.1\pm9.0$ & $-0.19\pm0.10$ &$1.95$ \\
 140  & $0.52\pm0.06$ & $-11\pm7.2$ & $-0.05\pm0.08$ &$2.78$ \\
      & $0.65\pm0.10$ & $-11\pm8.9$ & $-0.20\pm0.10$ &$1.95$\\
 240  & $0.54\pm0.07$ & $-13\pm7.4$ & $-0.04\pm0.08$ & $2.72$ \\
      & $0.65\pm0.10$ & $-11\pm8.9$ & $-0.19\pm0.10$ &$1.97$
\\ \hline\hline
\end{tabular}
\end{center}
\end{table}


Table ~\ref{tab3} lists the best fit parameters corresponding to the 
extracted WS for the SI. We observe that even 
in this case 
of diffuse emission from warm interstellar dust, there is a 
clear warp-like signature in all the four bands. Interestingly, the values 
of the phases are very similar to the stellar case. 
For both the SIs
we find that 
C and E bands show similar amplitudes within errors.
 However, for the D band we find that the $B_{med}$
  shows a larger amplitude of 
WS than $B_{int wt}$ and vice-versa for the A band. 
This is in contrast to the stellar case where the median
showed a larger amplitude of WS in all the 4 bands. For the case of 
SI of central tendencies, a better quality of fit to 
sinusoid is obtained for $B_{med}$ than $B_{int wt}$, for almost all 
the cases. 

\subsection{Latitude distribution of emission and warp signature from 
DIRBE data}
\begin {figure}
\begin {center}
\includegraphics[height=4.3cm]{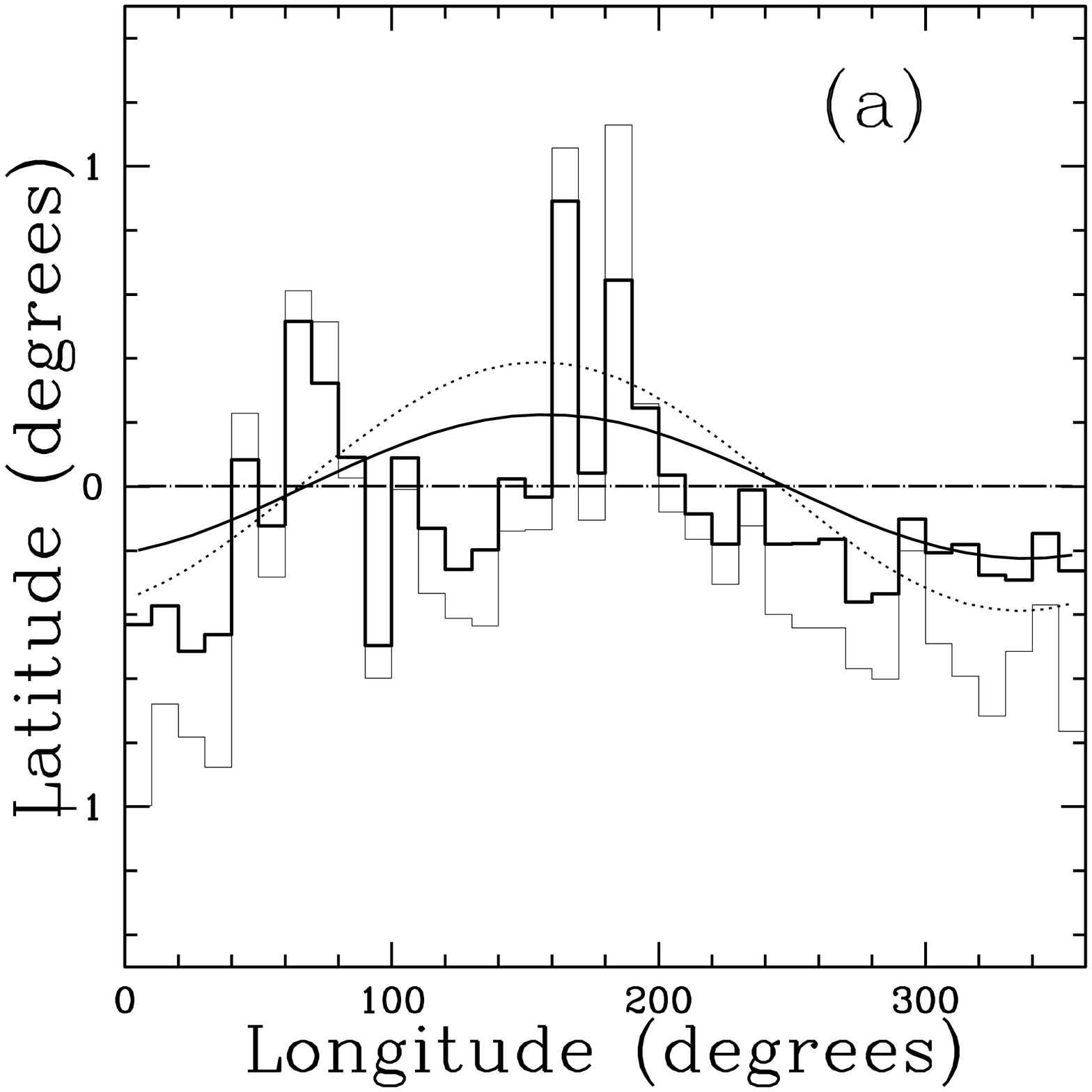}
\hskip 0.1cm
\includegraphics[height=4.3cm]{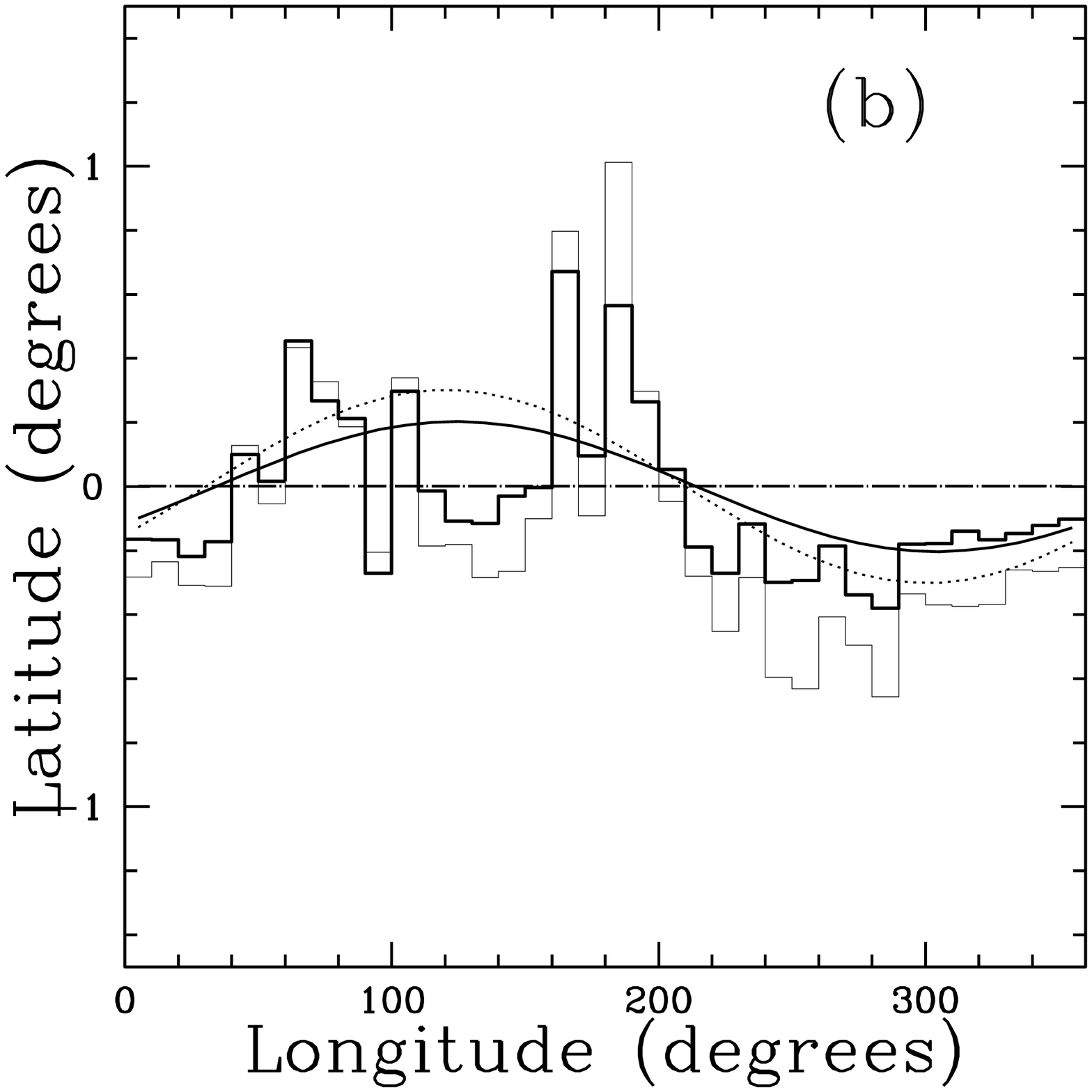}
\hskip 0.1cm
\includegraphics[height=4.3cm]{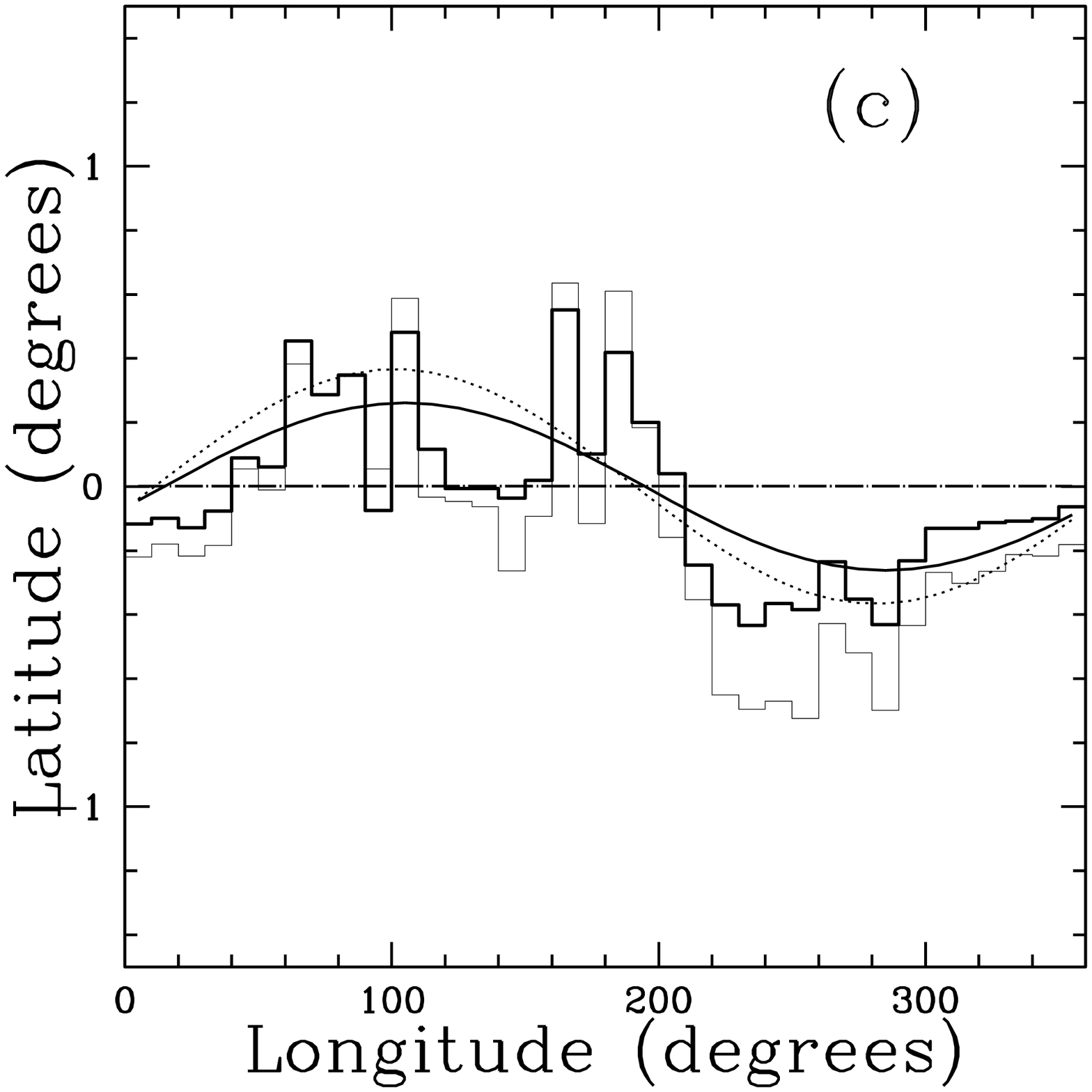}
\includegraphics[height=4.3cm]{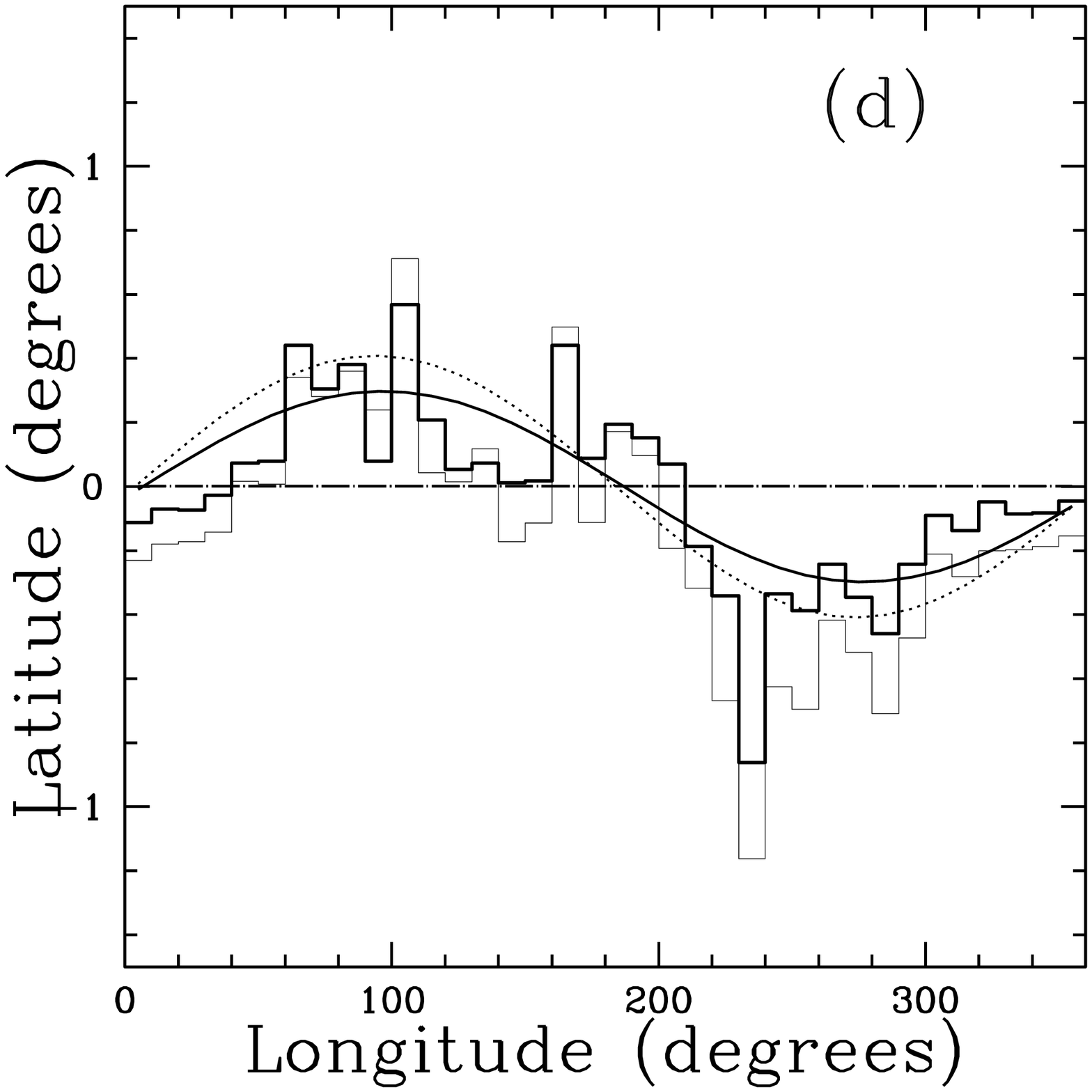}
\hskip 0.1cm
\includegraphics[height=4.3cm]{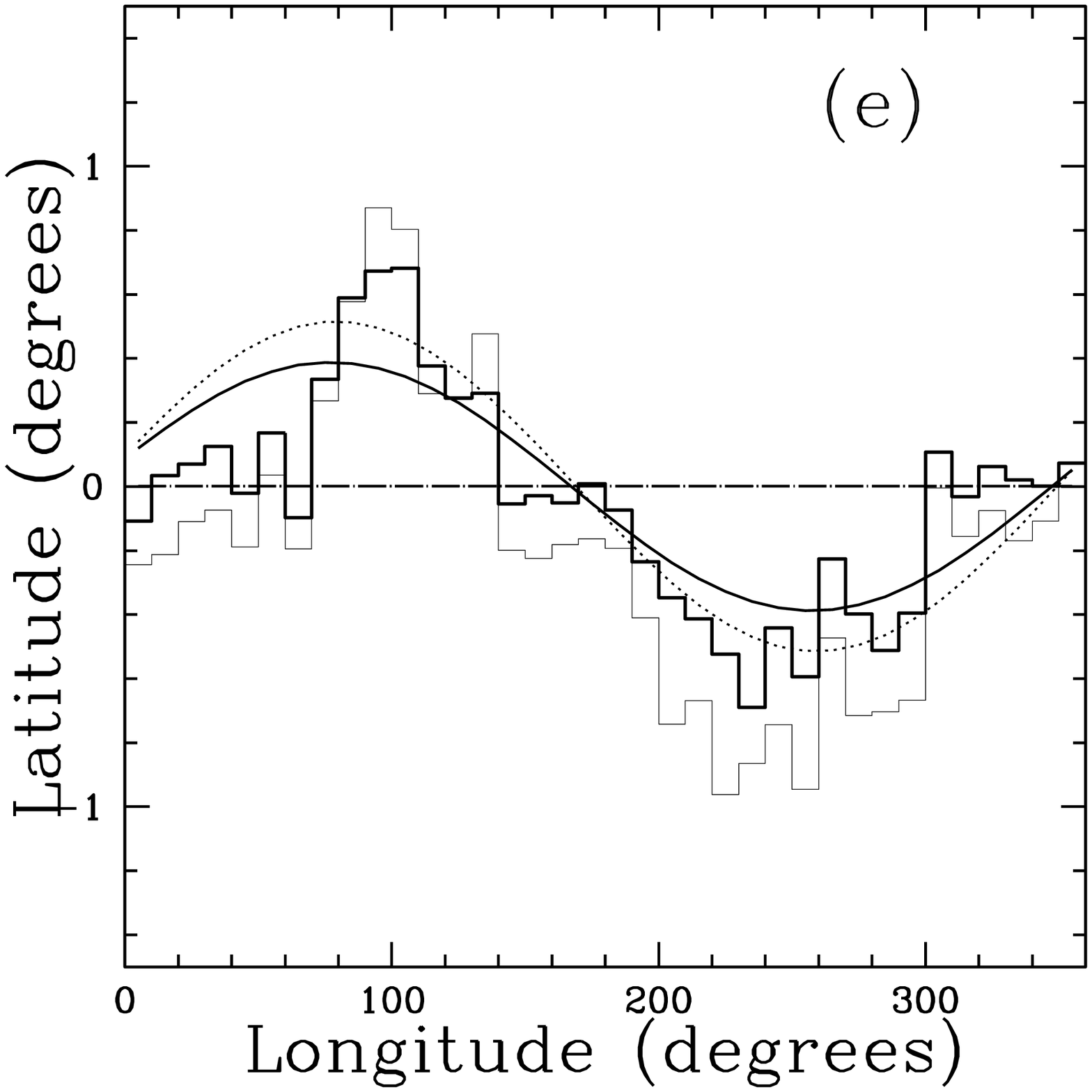}
\hskip 0.1cm
\includegraphics[height=4.3cm]{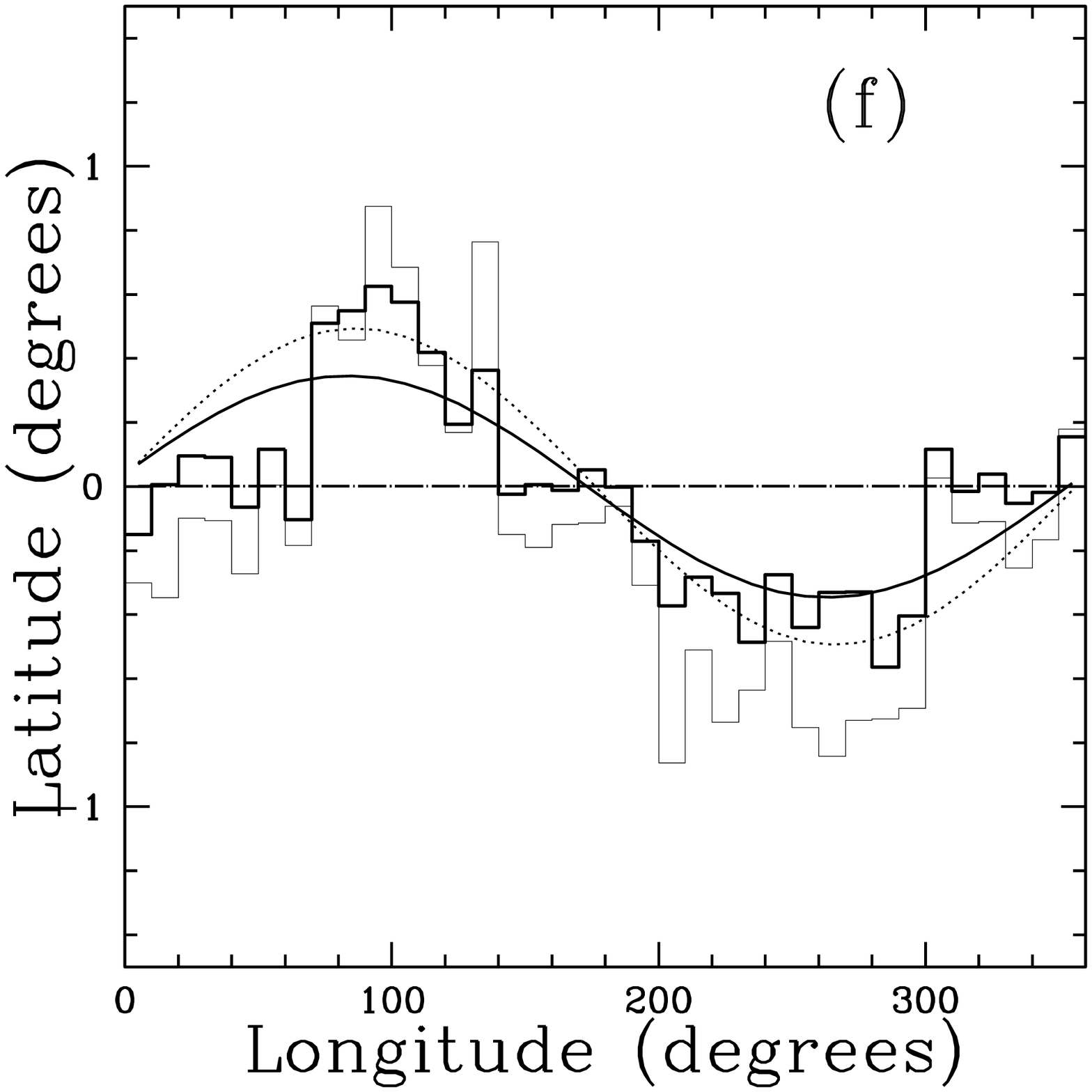}
\includegraphics[height=4.3cm]{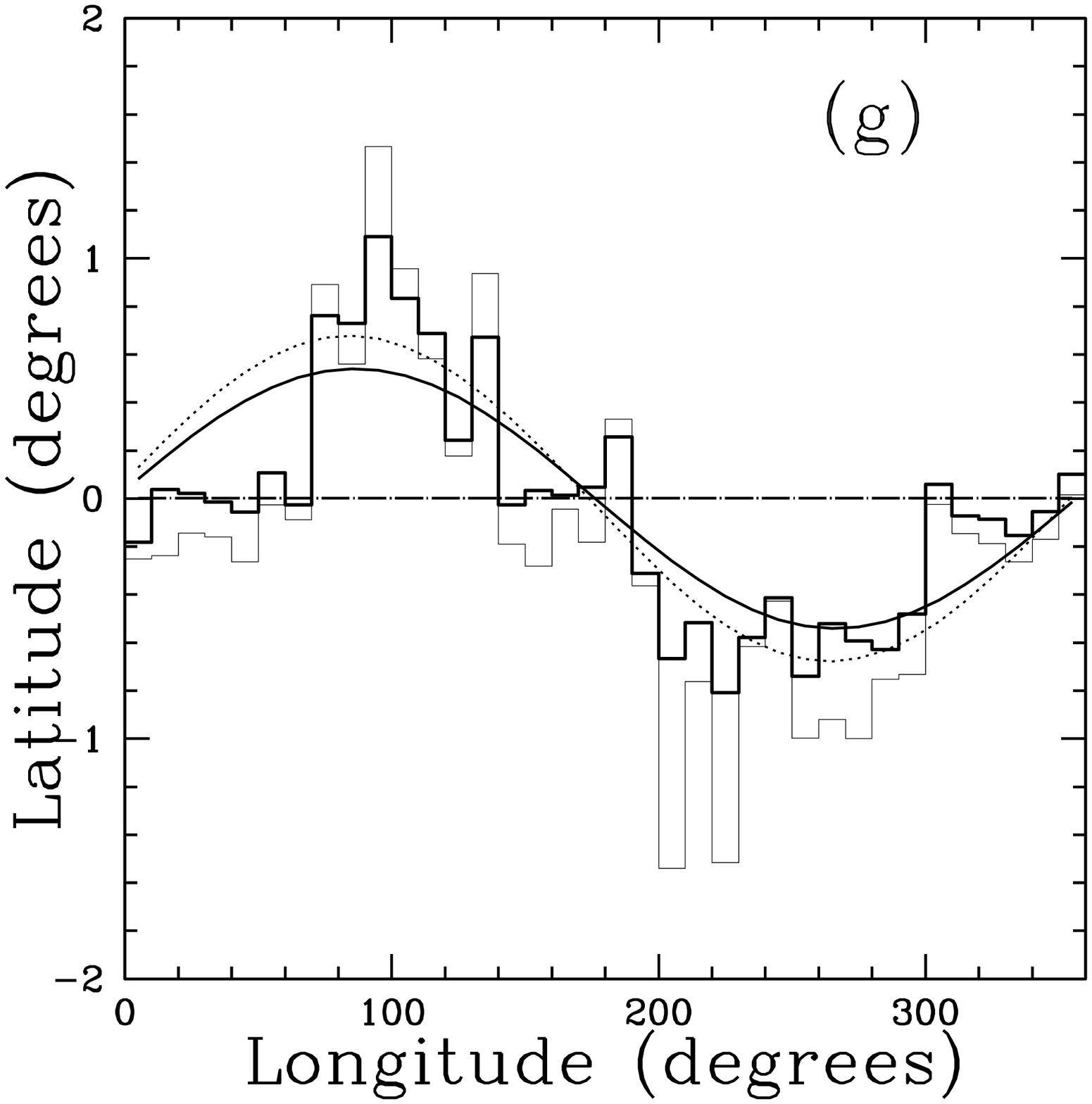}
\hskip 0.1cm
\includegraphics[height=4.3cm]{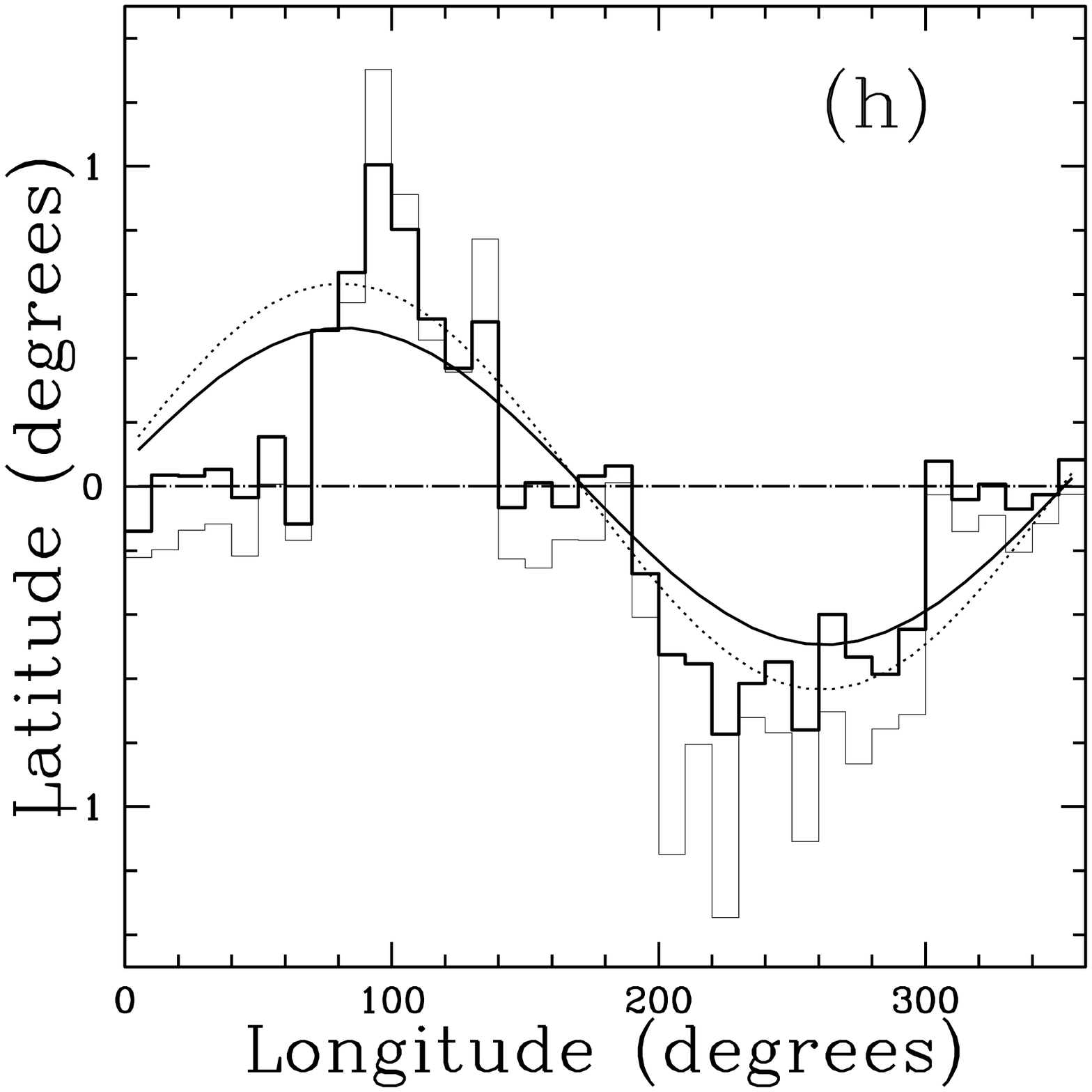}
\hskip 0.1cm
\includegraphics[height=4.3cm]{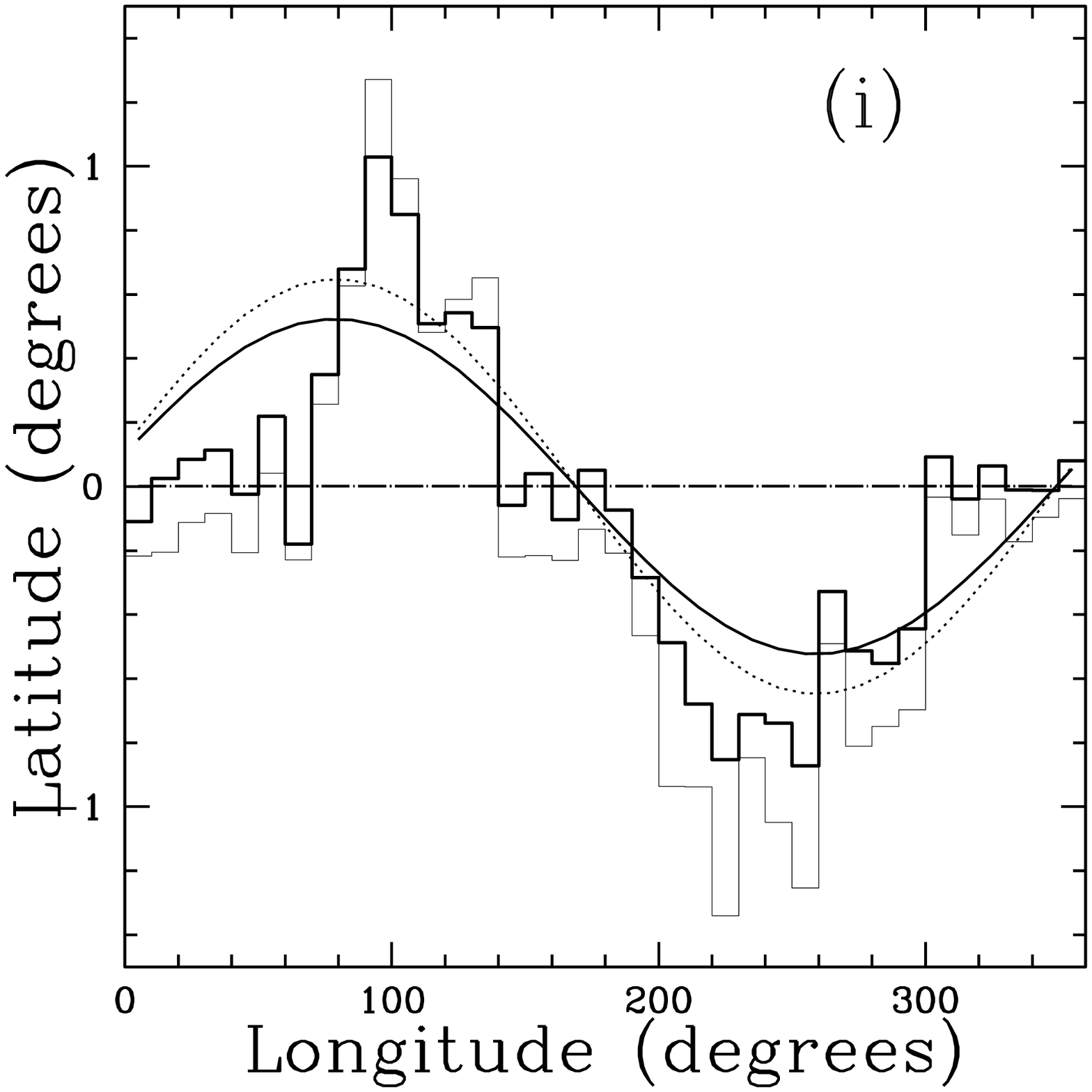}
\includegraphics[height=4.3cm]{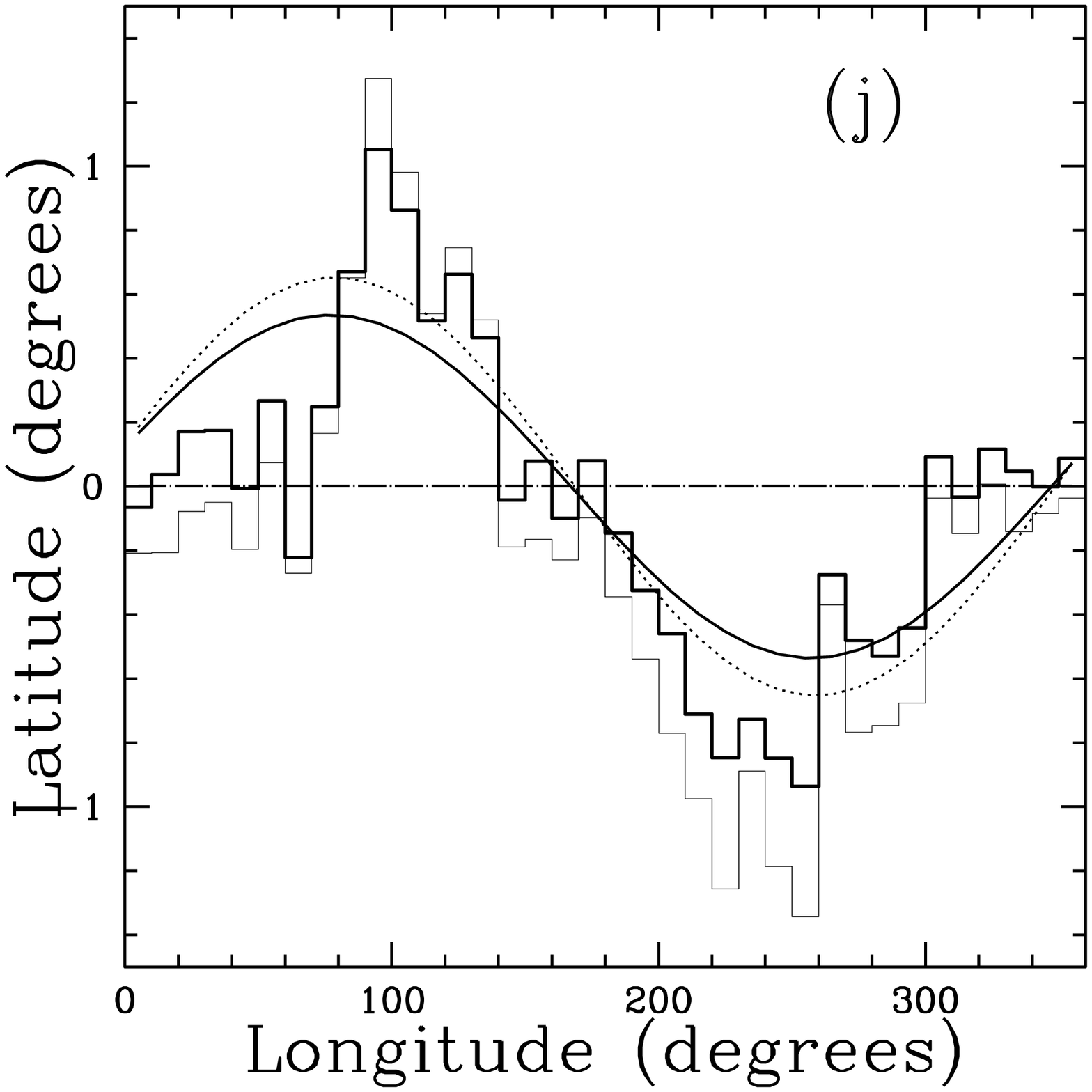}
\caption{Same as in Fig ~\ref{fig6} but for DIRBE intensity data in the ten bands: 
(a) 1.25 $\mu$m,(b) 2.2 $\mu$m, (c) 3.5 $\mu$m, (d) 4.9 $\mu$m, (e) 12 $\mu$m, 
(f) 25 $\mu$m, (g) 60 $\mu$m, (h) 100 $\mu$m, (i) 140 $\mu$m and (j) 240 
$\mu$m band. The solid and the dotted lines refer to the extracted warp 
signatures corresponding to the best fit sinusoid to $B_{intwt}$s and 
$B_{med}$s, respectively.
}
\vskip -0.6cm
\label{fig7}
\end{center}
\end {figure}

We have searched for WS from the emission of the
Galactic plane over the entire infrared waveband
using the DIRBE images following an analysis identical to that for
MSX images (Section 3.2.2). Despite poorer angular resolution, the large 
number of DIRBE wavebands (10) covering near (1.25 $\mu$m) to far (240 $\mu$m) 
infrared makes it a good candidate for comparison. The extracted
characteristic angular scales obtained from fitting latitude profiles (LP)
with exponential functions, $\gamma_{e}$, 
for the inner Galaxy have been listed in Table ~\ref{tab1}. 
The SIs for the Galactic midplane, $B_{intwt}$ \& $B_{med}$ 
for the entire longitude range, for each DIRBE band have been presented
in Figure ~\ref{fig7}. These SIs have also been searched for WS and the resulting
best fitting sinusoids have been overplotted on the same figure.

The WS has been detected in all the three (one LP and two SI) 
cases in each of the DIRBE bands. Even the phases of the WS,
as evident from Table ~\ref{tab2} (LP) (\& ~\ref{tab3}; SI), obtained for
9 (8) of the 10 bands are consistent with those obtained from MSX data
for the stars as well as the diffuse mid infrared emission.
It may be noted that at 1.25 $\mu$m, the only band where the WS phase (\& 
also the amplitude)
 deviates significantly from the other bands, the errors are the largest. 
This could be because the emission at 1.25 $\mu$m
originates mainly from the stars, which
would be affected most by extinction due to interstellar dust. 

\section{Characteristic angular and linear scale heights}
In the present study, we have determined the characteristic angular scales
of emission along the Galactic latitude,
which need to be translated to linear scales for drawing physical
inferences. For this purpose, we have carried out
calculations based on geometric description of the Galactic disk. 
The Galactic disk has been considered to be of radius 14 kpc 
and thickness 700 pc, with $R_{\odot} = 8$ kpc.
Our description neglects the effects of the Galactic bulge, the halo
as well as any warp feature.
The effect of these simplifications on our results is expected to
be negligible.

We need to estimate the integrated emission as a function of Galactic 
latitude for a given longitude, $I(b)$, with linear scale height
($z_h$) and radial scale length ($R_l$) as input parameters. 
It is reasonable to expect the observable intensity to scale with
the integral of the emission along the line of sight, $\rho$
(Hammersley et al. \cite{Ha95}). 
Hence, we write :
\begin{equation}         
I(b)\propto \int_{0}^{\rho_{max}(l,b)}e^{-\frac{\rho |sin(b)|}{z_h}-\frac{R}{R_l}}d\rho
\end{equation}
for a given $l$, where $R=\sqrt{R^2_{\odot} + (\rho \> cos \,b)^2 - 
2 \, R_{\odot}\> \rho \> cos\,b \,cos\,l}$
 is the galactocentric distance and $\rho_{max}(l,b)$ is the maximum length 
along a given line of sight which is only constrained by the 
finite size of the disk. 
These synthetic latitude profiles are found to be better represented by
exponential functions than gaussian as expected. 
We have fitted exponential 
functions of the form $\rho_oe^{-\frac{|b|}{\psi_h}}+\eta$ 
to these latitude profiles,
with $\rho_o$, $\psi_h$ and $\eta$ as free parameters. The
$\psi_h$ corresponds to characteristic angular scale and $\eta$ 
represents the background (no latitude dependent background is anticipated due 
to the symmetry of the geometry). 
The study has been carried out at two 
representative longitudes $30^{\circ}$ and $90^{\circ}$. 
The lines of sight closer towards the Galactic centre 
have been avoided due to difficulty in comparing the calculations
with measurements, since our description neglects the more 
complex details (like the bulge) which could be important.
 The following values of $R_l$ : 1.5, 2.0, 2.5, ..., 7.0, 7.5 
and $\infty$ kpc
($R_l=\infty$ corresponds to constant radial density distribution) 
have been explored. 
The set of 
$z_h$ values explored are 40, 50, 60, ..., 90, 100, 125, 150, ..., 275, 300 pc.

\begin {figure}
\begin {center}
\includegraphics[height=6.0cm]{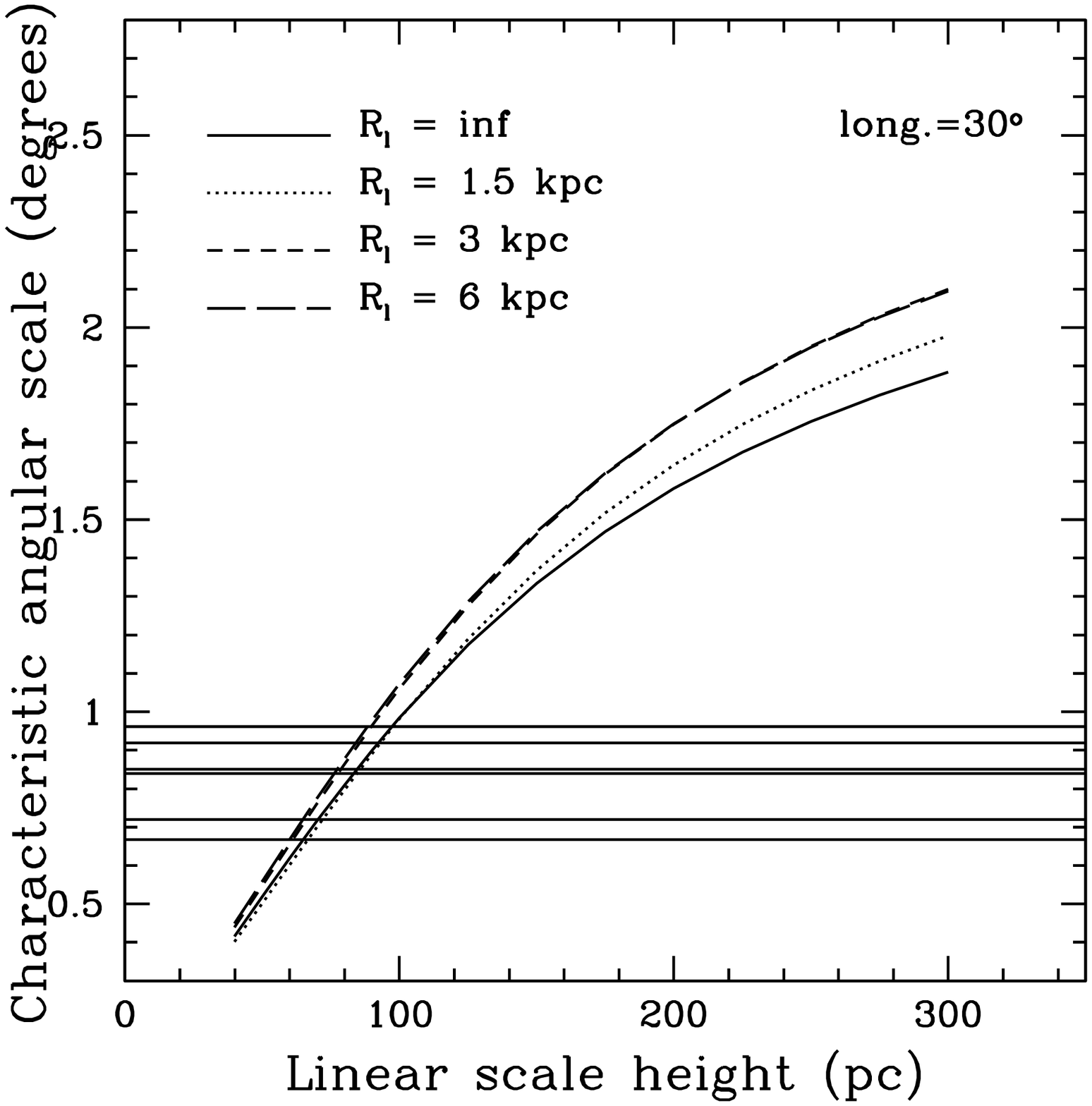}
\hskip 0.5cm
\includegraphics[height=6.0cm]{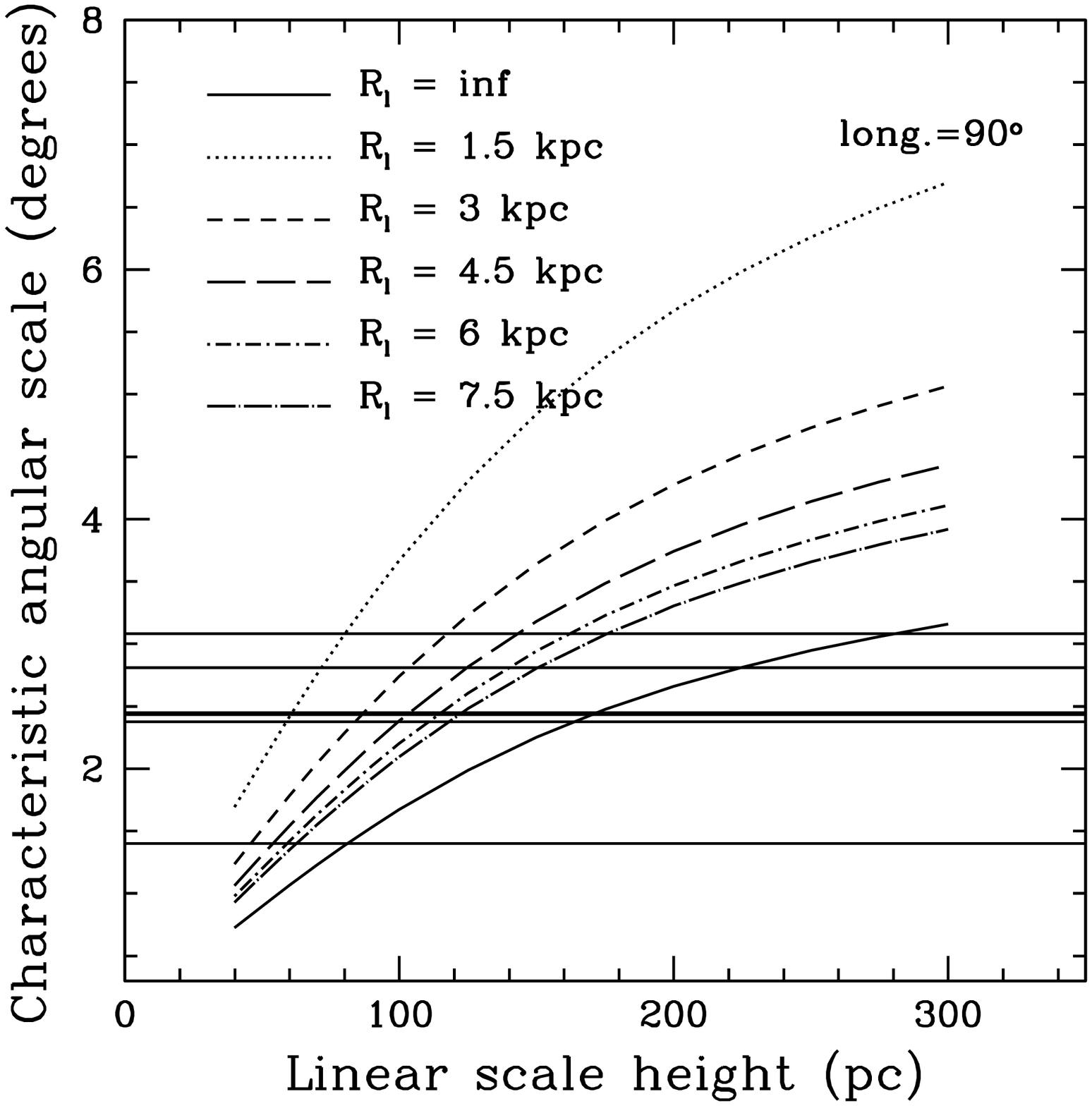}
\caption{Relations between the linear ($z_h$) and the angular ($\gamma_e$) 
scale heights from a geometric description of the Galactic disk
(for a heliocentric observer), 
for different assumed radial scale lengths, R$_l$. The two panels 
represent longitudes $l=30^{\circ}$ \& $l=90^{\circ}$. 
The $\gamma_e$s obtained from the exponential fits for 
the diffuse mid infrared emission in 4 MSX bands 
(for A, C, D \& E bands : 0.85$^{\circ}$, 0.92$^{\circ}$, 
0.96$^{\circ}$ \& 0.67$^{\circ}$ for $l$=30$^{\circ}$; 
and 3.08$^{\circ}$, 2.45$^{\circ}$, 
 2.43$^{\circ}$ \& 2.37$^{\circ}$ for $l$=90$^{\circ}$) 
and the two DIRBE bands
(for 12 \& 25 $\mu$m bands : 0.84$^{\circ}$, 
 0.72$^{\circ}$ for $l$=30$^{\circ}$;
 and 2.81$^{\circ}$, 1.40$^{\circ}$ for $l$=90$^{\circ}$) 
are also plotted as horizontal solid
 lines for comparison.
}
\label{fig8}
\end{center}
\end {figure}

The results of our calculations (for $l$=30$^{\circ}$ and 90$^{\circ}$) 
connecting the 
characteristic angular scale ($\gamma_e$) with linear scale height ($z_h$), 
for selected values of the radial 
scale lengths ($R_l$), have been presented in Figure ~\ref{fig8}. 
The larger separation among curves corresponding
to different values of $R_l$ for $l$=90$^{\circ}$ 
as compared to $l$=30$^{\circ}$,
is entirely expected from the shape and size of the Galactic disk
and the heliocentric location of the observer.
Overplotted as horizontal lines in this figure, 
are the values of $\gamma_e$ for mid infrared wavebands at
corresponding longitudes,
as obtained earlier 
from fitting the observed latitude profiles (Section 3.2.2 and 3.3) for the 
4 MSX bands and 2 DIRBE bands (12 and 25 $\mu$m).
This comparison helps constraining the values of $R_l$ and $z_h$ 
for the distribution of warm interstellar dust. 
From the overlap of the observed and the model curves for $l=30^{\circ}$, 
we infer that the preferred range of $z_h$, is $\sim60-100$ pc, 
while $R_l$ cannot be constrained due to the 
insensitivity to this parameter. However, the situation at $l=90^{\circ}$ is 
quite favourable and provides a handle for $R_l$. 
Assuming $z_h$ to be in the above range, 
it is evident that the radial scale
length, $R_l$, should be less than $\sim$ 6 kpc.
Although the value of $\gamma_e$ obtained from DIRBE 25 $\mu$m band,
1.4$^{\circ}$, 
seems incompatible with the above (Fig. 8, $l=90^{\circ}$), 
we note that the values corresponding to the immediate neighbouring
longitude bins are systematically higher (2.0$^{\circ}$ \& 2.2$^{\circ}$),
consistent with the above constraint on $R_l$.
Giving equal weightage to the six mid-IR bands,
we find the most favoured value of $R_l$ to be 2.81 $\pm$ 0.63 kpc
corresponding to $z_h$ = 80 pc 
(central value of our implied range).
Similarly, for the far infrared bands representing the cold interstellar
dust, the scale heights $z_h$ have been found to be $\sim 60$ pc.

\begin {figure}
\begin {center}
\includegraphics[height=9.0cm]{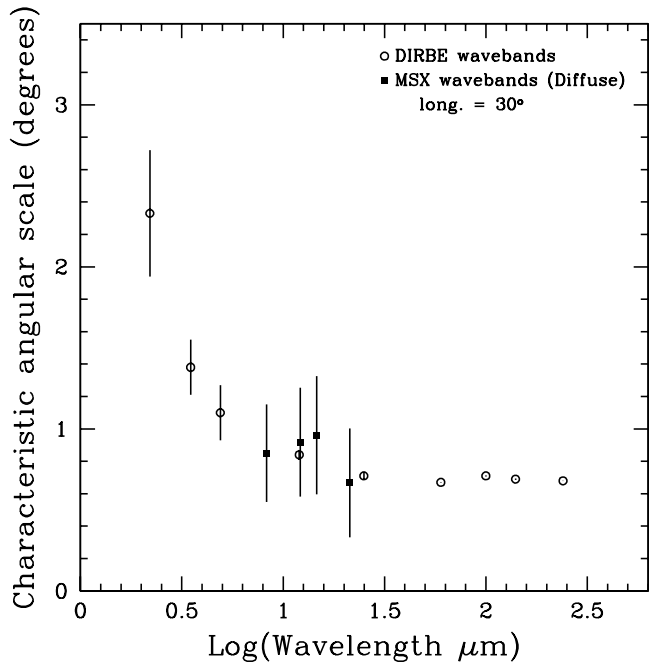}
\caption{Wavelength dependence of the angular scale height, $\gamma_e$, 
for the longitude bin $l$=30$^{\circ}$. The open circles represent the DIRBE 
wavebands for $\lambda \ge 2.2$ $\mu$m and the solid squares represent the 
diffuse emission in the MSX bands at 8, 12, 14 and 21 $\mu$m. 
The error bars indicate 1$\sigma$ uncertainity.
At longer wavelengths the errors are smaller than the symbol size.
}
\vskip -0.6cm
\label{fig9}
\end{center}
\end {figure}

In addition, it is instructive to compare the characteristic angular scales,
$\gamma_e$, 
 obtained in the present study as a function of wavelength. For this, we have
selected a moderate longitude bin ($l=30^{\circ}$) 
where the complexities due to
additional components of the Galactic structure near the Galactic center is 
absent, yet the signal to noise ratio is reasonably good for reliable
determination of $\gamma_e$s.
The variation of $\gamma_e$ with wavelength, at l=$30^{\circ}$,
corresponding to the DIRBE and MSX wavebands ($\lambda \ge$ 2.2 $\mu$m)
have been presented in Figure ~\ref{fig9}.
We find that $\gamma_e$ drops with wavelength from 
near to mid infrared and remains more or less constant (within errors)
thereafter upto far infrared wavelengths.

While the near infrared emission is expected to be dominated by  
stars, the mid infrared bands  
sample the warm interstellar dust in the vicinity of star forming regions. 
On the other hand, the far infrared emission is predominantly from cold 
dust heated by average interstellar radiation field. 
The observed decrease in $\gamma_e$ obtained from 2.2 to 8.3 $\mu$m bands 
could possibly be linked to the extinction due to dust. 
Typical wavelength dependence of dust extinction being
$\propto \lambda^{-\alpha}$ (1 $< \alpha <$ 2),
the extinction is expected to be much larger for the near 
infrared as compared to mid or far infrared wavebands. 
The effect of  
extinction is largest on the Galactic plane ($b$=0) and 
progressively reduces at higher latitudes.
 The observed peak emission being reduced due to attenuation, the fit to 
the data could indicate an effectively larger angular scale height.

\section{Comparison with other results}
In this section we compare our results about the structure and WS of the 
Galactic disk with other studies of the distribution of stellar and the 
interstellar components.

\subsection{Indicators of structure of the Galactic disk}
The present study has constrained the scale height and radial
scale length for the distribution of warm interstellar dust
($z_h=60-100$ pc, $R_l < 6$ kpc; Section 4).
Our $z_h$ is close to that for the ISM (140 pc) from the successful model
of the Galactic structure by Robin et al. (\cite{Ro03}), although their $R_l$
is higher (4.5 kpc).
The distribution of dust studied by Freudenreich (\cite{Fr98}) using 
near infrared measurements imply : $z_h \sim 150 - 200$ pc \&
$ R_l \sim 3.0 - 3.3$ kpc,
 which is in reasonable agreement with our $R_l$. 
Drimmel \& Spergel (\cite{Dr01}), have obtained the values $z_h$=134 pc \& 
$R_l$=2.26 kpc for the cold dust component, using 
DIRBE 240 $\mu$m data, which compares very well with our results.   

\subsection{Signatures of Galactic warp}
In this subsection, our results of WS have been compared with other studies 
for the stellar, the warm and cold interstellar dust components. The 
Table ~\ref{tab4} presents the summary of this overall comparison.

\subsubsection{Warp signature from the stellar component}
The WS for the stellar component obtained here using 
the MSX PSC, is in very good agreement with
that obtained by Djorgovski \& Sosin (\cite{Dj89}) from the distribution 
of IRAS point sources, despite their sample covering
a much larger latitude range of $|$b$| \leq 10^{\circ}$.
The estimates for the WS phase, $\phi$, obtained here from
both the approaches (using 
latitude profiles for the inner Galactic plane, LP; \& using
statistical indicators for 
the entire Galactic plane, SI) match extremely well
with the IRAS results (see Table ~\ref{tab4}).
The WS amplitude obtained for the SI case is also consistent with 
the IRAS result, although the LP case (representing only the inner Galaxy)
implies somewhat larger value.  
Similar consistency is also observed for the results obtained here
using DIRBE data in L (3.5 $\mu$m) \& M (4.9 $\mu$m) bands
for both LP \& SI.
Lopez-Corredoira et al. (\cite{Lo02b}) have analysed the data from Two 
Micron All Sky Survey (2MASS) in the context of Galactic warp.
Their best fit warped model 
(Fig. 15, $b$ = +3$^{\circ}$ / $b$ = -3$^{\circ}$) 
corresponds to a value of +2$^{\circ}$ for our phase 
parameter $\phi$, which is 
again similar to those obtained here from all the four 
bands of MSX PSC as well as L \& M bands of DIRBE.
Freudenreich (\cite{Fr98}) from a study of the Galactic disk
obtain -0.08$^{\circ}$ $< \phi <$ 0.8$^{\circ}$,
 which again is consistent with our results.

 It is interesting to note that,
stellar distributions obtained from MSX and IRAS
 show fair amount of symmetry between the positive 
and negative peaks of the warp
signature, unlike the case for H I distribution (Burton \cite{Bu88}).


\begin{table*}
\begin{center}
\caption{
Comparison of parameters of warp signature (amplitude and phase)
 for the Galactic disk
 obtained from different studies.}
\label{tab4}
\begin{tabular}{|c c c c c c |} \hline\hline
 $\lambda$($\mu$m)  & Survey & Comp. & LP/SI$^a$ & WS Amplitude & WS Phase\\ 
 (Band) & & & & (deg) & (deg) \\ \hline

 2.2$^b$     & 2MASS & Stellar & ... & ... & 2 \\
(NIR)& & & & & \\
&&&&& \\

2.2$^c$  & DIRBE & Stellar & LP & 0.5 & 19  \\
(NIR)  &  &  &  SI & 0.2 - 0.3 & 30 - 34 \\
&&&&& \\

 3.5, 4.9$^c$ & DIRBE & Stellar & LP &0.3 -- 0.4 &  15 \\
(NIR)   &  &  &  SI &0.3 -- 0.4 & 4 -- 15  \\
&&&&& \\

 8.3--21$^c$ & MSX & Stellar & LP &0.4 -- 0.5 & 3 -- 7 \\
(MIR)  &  &  & SI &0.2 -- 0.4 & 1 -- 18  \\
&&&&& \\

12--60$^d$  & IRAS & Stellar & ... &0.2 -- 0.3 &0 -- 10  \\ 
(MIR/FIR) & & & & & \\ \hline
&&&&& \\

8.3$^c$  & MSX & Warm dust & LP &0.4 & 18  \\
(MIR)  &  &  & SI &0.2 -- 0.4 & -3 -- 2 \\
&&&&& \\

12--21$^c$ & MSX & Warm dust & LP &0.4 -- 0.7 & -6 -- 11  \\
(MIR)   &  &  & SI &0.4 -- 1 & -25 -- -12  \\
&&&&& \\

 12, 25$^c$ & DIRBE & Warm dust & LP & 0.4 & 11 -- 18  \\
(MIR)    &  &  & SI &0.4 -- 0.5 & -13 -- -4 \\
&&&&& \\

 60--240$^c$ & DIRBE & Cold dust & LP &0.4 -- 0.5 & 12 -- 16  \\
(FIR)    &  &  & SI &0.5 -- 0.7 & -13 -- -4  \\
&&&&& \\ \hline \hline

\end{tabular}
\end{center}
$^a$ From Latitude Profiles, -90$^{\circ}<l\le$90$^{\circ}$ (LP)
/ Statistical Indicators, 0$^{\circ}<l\le$ 360$^{\circ}$(SI) \\
$^b$ Lopez-Corredoira et al. (\cite{Lo02b}) \\
$^c$ Present work \\
$^d$ Djorgovski \& Sosin (\cite{Dj89}) \\
\end{table*}


\subsubsection{Warp signature from the interstellar dust component}
The diffuse mid infrared ($8-21 \mu$m) emission detected by MSX is expected
to originate from the warmer (T $\ga$ 100 K) component of the
interstellar dust. In order to be heated to such high temperatures, the
dust grains need strong local radiation fields, e.g. 
 near young stellar objects (OB stars). 
Hence, detection of WS from diffuse mid infrared emission 
provides information about the distribution of warm
dust around luminous young stars in the Galactic plane.
Our results of the WS for the diffuse component 
obtained from the 4 bands of MSX and the two mid infrared bands
of DIRBE (12 \& 25 $\mu$m) have also been compared in Table ~\ref{tab4}. 
The $\phi$s obtained are consistent within errors.
The amplitudes inferred from the MSX and DIRBE are similar with the former
being slightly larger.
An identical systematic shift ($\sim$ 20$^{\circ}$) 
in $\phi$ can be noticed between the values
obtained for the inner Galactic plane (from LP)
and those for the entire disk (from SI), for all the mid infrared bands
of MSX as well as DIRBE.

The far infrared emission originates from the colder 
(T $\la$ 30 K) component of interstellar
dust in the Galactic plane. The WS detected
in 60--240 $\mu$m wavebands of DIRBE  
should correspond to the distribution 
of this component. The $\phi$s 
agree very well with those for the stellar as 
well as the  warm dust component (Table ~\ref{tab4}). The amplitude
of the WS is similar to that obtained for the warmer dust.

\section{Summary}
 Taking advantage of the recent high resolution ($\sim$ 18\arcsec)
 Galactic plane survey in the mid infrared (8, 12, 14 \& 21 $\mu$m)
 by the Midcourse Space Experiment (MSX), the distribution of stars
 (from the Point Source Catalog) as well as warm interstellar dust
 (from the panoramic images) in the Galactic disk have been
 investigated. The location of the Galactic mid-plane and the
 characteristic angular scales along latitude have been obtained
 from latitude profiles ($-90^{\circ} < l \le 90^{\circ}$), and statistical
 indicators (entire Galactic plane). Using a simple geometric
 description of the Galactic disk, the angular scales have been
 translated to scale heights ($z_h$) and radial length scales ($R_l$).
 The following constraints have been obtained for the distribution
 of warm (T $\ga$ 100 K) interstellar dust in the Galactic disk :
 $ R_l < 6$ kpc and $60 \la z_h \la 100$ pc. 
These are
 consistent with results from other studies.

   Signatures of warp-like feature in the distribution of stars
 as well as warm interstellar dust have been searched for using
 the inferred locations of the Galactic mid-plane as a function
 of longitude. The warp signature (WS) has been detected in all the
 cases : for point sources as well as diffuse emission 
in
 each of the 4 MSX bands. The WS for each case has been quantified
 by two parameters depicting its amplitude and phase (longitude
 of the line of nodes).

 Our analysis/study  of the mid infrared emission has been
 extended to cover the entire infrared waveband by using the data
 from the DIRBE/COBE experiment (10 bands; 1.25--240 $\mu$m).
 Despite limited angular resolution of DIRBE, signatures of warp
 have been detected in each band with very  similar phase as found
 from the MSX data. The inferred amplitudes of WS lie in the
 range : $0.2^{\circ}-0.7^{\circ}$ (for $\lambda \ge 2.2 \mu$m) in latitude.
 Results from the present study have also been compared with those
 from other similar studies based on the IRAS and 2MASS surveys.
 
\begin{acknowledgements}
It is a pleasure to thank the referees -- M. Lopez-Corredoira
and the anonymous referee -- for comments 
 and suggestions, which greatly improved the  scientific content of the paper.
We thank Sean Carey and Michael Egan for providing us with MSX 
panoramic mosaics. This research made use of data products from the Midcourse 
Space Experiment, the processing of which was funded by the Ballistic Missile 
Defense Organization with additional support from NASA Office of Space Science.
 This research has also made use of the NASA/IPAC Infrared Science Archive, 
which is operated by the Jet Propulsion Laboratory, California Institute of 
Technology, under contract with National Aeronautics and Space Administration.

The COBE datasets were developed by the NASA Goddard Space Flight Center under
 the guidance of the COBE Science Working Group. We acknowledge the use of the 
Legacy Archive for Microwave Background Data Analysis (LAMBDA). Support for 
LAMBDA is provided by the NASA Office of Space Science.
We thank Annie Robin for letting us use her model of 
stellar population synthesis and for helpful discussions. 
S.V. acknowledges the TIFR Endowment Fund for partial 
financial support.
\end{acknowledgements}




\end{document}